\documentclass[10pt,
               twocolumn,
               showpacs,
               superscriptaddress,
               floatfix,
               longbibliography,
               aps,
               prl]{revtex4-2}

\usepackage[english]{babel}

\usepackage{graphicx,wrapfig}
\graphicspath{{./}}

\usepackage{amsmath,amssymb,bm,physics,upgreek,cancel,extarrows}
\usepackage[colorlinks=true,
            allcolors=blue]{hyperref}

\usepackage{acro}
\DeclareAcronym{rf}{
    short=RF,
    long=radio frequency,
}
\DeclareAcronym{fmo}{
    short=FMO,
    long=Fenna–Matthews–Olson,
}
\DeclareAcronym{sm}{
    short=SM,
    long=Supplemental Material,
}
\DeclareAcronym{em}{
    short=EM,
    long=End Matter,
}

\usepackage{comment}
\usepackage{subfiles}
\usepackage{lipsum}

\newcommand{\ee}{\mathrm{e}}
\newcommand{\ii}{\mathrm{i}}
\newcommand{\pii}{\uppi}
\newcommand{\ed}{\downarrow}
\newcommand{\eu}{\uparrow}
\newcommand{\nuRF}{\nu_\mathrm{RF}}
\newcommand{\omgRF}{\omega_\mathrm{RF}}
\newcommand{\pe}{\bm{p}_\mathrm{e}}
\newcommand{\pen}{\bm{p}_\mathrm{en}}
\newcommand{\re}{\bm{r}_\mathrm{e}}
\newcommand{\ren}{\bm{r}_\mathrm{en}}
\newcommand{\me}{m_\mathrm{e}}
\newcommand{\men}{m_\mathrm{en}}
\newcommand{\vepsv}{\varepsilon_\mathrm{v}}
\newcommand{\vepse}{\varepsilon_\mathrm{e}}
\newcommand{\ecz}{E_{\mathrm{c},z}}
\newcommand{\nexc}{{N_\mathrm{exc}}}
\newcommand{\aB}{a_\mathrm{B}}
\newcommand{\muB}{\mu_\mathrm{B}}
\newcommand{\mud}{\mu_\mathrm{d}}
\newcommand{\muq}{\mu_\mathrm{q}}
\newcommand{\xhoz}{x_{\mathrm{ho},z}}
\newcommand{\omgc}{\omega_\mathrm{c}}
\renewcommand{\cap}{$^{40}\mathrm{Ca}^+$}
\newcommand{\srp}{$^{88}\mathrm{Sr}^+$ }
\newcommand{\bap}{$^{138}\mathrm{Ba}^+$}
\newcommand{\elgf}{g_\mathrm{e}}
\newcommand{\lande}{g_\mathrm{L}}

\begin{document}

%%% FIGURES %%%%%%%%%%%%%%%%%%%%%%%%%%%%%%%%%%%%%%%%%%%%%%%%%%%%%%%%%%%%%%%%%%%%
\def\figMainPicture{
\begin{figure*}[t]
    \centering
    \includegraphics{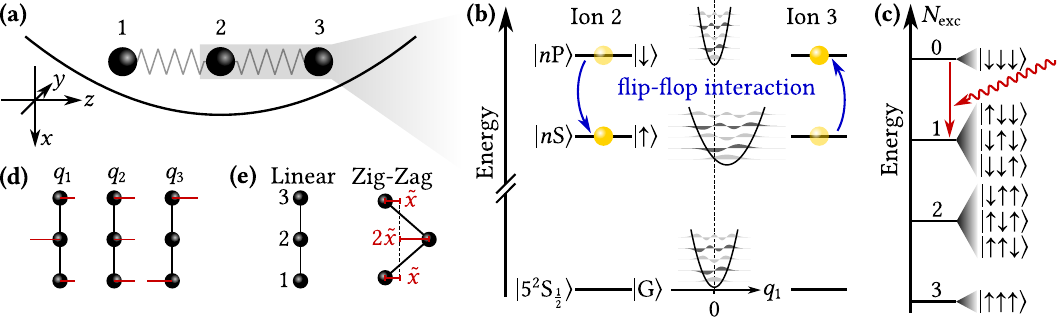}
    \caption{
        \textbf{Elementary degrees of freedom of the quantum simulator}. (a) Ion crystal with three ions, labelled 1, 2, 3, confined in a linear Paul trap's  potential \cite{leibfried2003}, represented by the black parabola. (b) Electronic structure of the interacting ions 2 and 3, depicted in panel (a). Here, $\ket*{\mathrm{G}}$ is the electronic ground state and $\ket*{n\mathrm{S}}$, $\ket*{n\mathrm{P}}$ are Rydberg S and P states. Blue arrows indicate resonant `flip-flop' processes due to the dipole-dipole interaction of Rydberg ions. The trapping potential experienced by the ions depends on the electronic state. This is illustrated by the sketches of the potentials associated with vibrational mode $q_1$ [see panel (d)], where the vibrational eigenstates are shown in grey. (c) For three ions, the many-body electronic state space splits into the four resonant subspaces, labelled by the number of excitons $\nexc$. The red arrows depict an incoming photon creating an exciton, which can be transported by the dipolar `flip-flop' processes shown in panel (b). (d) Displacement signatures of the three considered collective modes $q_1$, $q_2$, and $q_3$. (e) Different conformations of the ion crystal \cite{fishman2008}. At equilibrium, an ion crystal consisting of three ions can display a linear or a zig-zag geometry. Which geometry is selected can be controlled with the trap parameters \cite{schmidtkaler2011}. In the zig-zag regime, the geometry of the ion crystal is quantified by the reaction coordinate $\tilde{x}$. It measures the displacement of ions 1 and 3 from the trap axis, which, due to symmetry, is half the displacement of ion 2.
    }
    \label{fig:MainPicture}
\end{figure*}
}
\def\figBlockage{
\begin{figure}[t]
    \centering
    \includegraphics{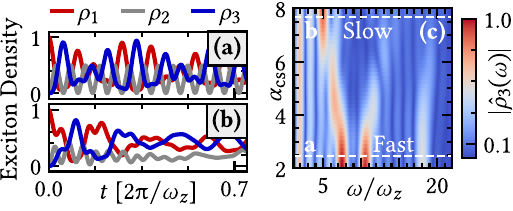}
    \caption{
        \textbf{Exciton transport slowed down by molecular vibrations.} (a) Time evolution of exciton densities $\rho_k$ governed by Hamiltonian \eqref{eq:ThreeIonModel}. The collective mode $q_1$ absorbs the momentum of the photon, such that it undergoes oscillations described by the motional coherent state $\ket*{\alpha_\mathrm{cs}}$; the modes $q_2$ and $q_3$ are initially in the vacuum. The oscillations of the blue and red curves show that the exciton is transported back and forth between ions 1 and 3. (b) Analogous dynamics to panel (a), but for a larger initial `kick' amplitude $\alpha_\mathrm{cs}$. Here, the transport is slower than in panel (a). (c) Frequency spectrum of exciton dynamics. On the colour axis we show the Fourier transform of the exciton density $\rho_3(t)$ at ion 3, $\hat{\rho}_3(\omega)$, for a range of amplitudes $\alpha_\mathrm{cs}$. Towards larger amplitudes $\alpha_\mathrm{cs}$ the high-frequency peaks become weaker and only one low-frequency peak remains. The spectra labelled `Fast' and `Slow' correspond to the dynamics shown in panels (a) and (b), respectively. To enhance visibility, the spectrum is normalised to the maximal value in the depicted area and it is only shown for $\omega\slash\omega_z\geq0.9$, to cut off the peak at $\omega=0$ due to the constant offset in the strictly positive exciton density $\rho_3(t)$. The parameters for the simulation are $n=45$ (principal quantum number of $^{88}\mathrm{Sr}^+$), $\alpha=10^9\,\mathrm{Vm^{-2}}$, $\beta=0.5\times10^7\,\mathrm{Vm^{-2}}$, $\nuRF=16\,\mathrm{MHz}$, and $\Delta^{(2)}=\Delta^{(3)}=0$, leading to $J=-6.4$, $\varepsilon_\eu=-0.32$, $\varepsilon_\ed=1.06$, and $\omega_z=2\pii\times0.53\,\mathrm{MHz}$.
    }
    \label{fig:Blockage}
\end{figure}
}
\def\figVAET{
\begin{figure}[t]
    \centering
    \includegraphics{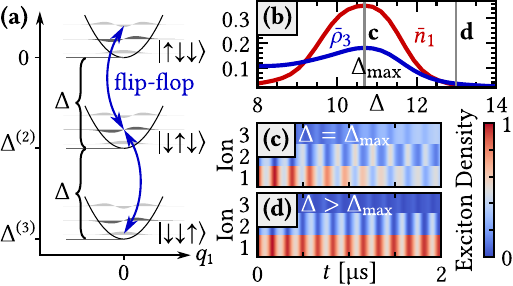}
    \caption{
        \textbf{Vibrationally assisted exciton transfer.} (a) Scenario for vibrationally assisted exciton  transport. The black parabolas sketch the molecular potentials associated with each electronic state in the single-excitation manifold, as functions of the collective coordinate $q_1$. Their energy offsets are controlled by the detunings in Hamiltonian \eqref{eq:ThreeIonModel}: we set $\Delta^{(2)}=-\Delta$ and $\Delta^{(3)}=-2\Delta$. For particular values of the energy offset $\Delta$, the interplay between flip-flop processes and vibrational excitation leads to a facilitation of the exciton transport. Blue arrows indicate effective `flip-flop' processes which alter the motional state (shown in grey). (b) Time-averaged exciton density at ion 3, $\bar{\rho}_3$, and time-averaged phonon occupation number, $\bar{n}_1$, in the mode $q_1$, for a range of energy offsets $\Delta$. Both quantities are averaged over an interval of 2\,\textmu{s}. The grey lines mark two different values of $\Delta$ for which we show the spatially resolved transport in (c) and (d). (c) Time evolution of the exciton densities $\rho_k$ at the maximum $\Delta=\Delta_\mathrm{max}$ in (b). Initially, the exciton is localised on ion 1 and the vibrational degrees of freedom are in the vacuum. (d) Analogous dynamics to (c) but for $\Delta>\Delta_\mathrm{max}$. For the simulation, $n=50$ (principal quantum number of $^{88}\mathrm{Sr}^+$), $\alpha=10^9\,\mathrm{Vm^{-2}}$, $\beta=0.15\times10^7\,\mathrm{Vm^{-2}}$, and $\nuRF=15\,\mathrm{MHz}$, leading to $J=-5.5$, $\varepsilon_\eu=-2.3$, $\varepsilon_\ed=7.6$.
        }
        \label{fig:VAET}
\end{figure}
}
\def\figRelaxation{
\begin{figure}[t]
    \centering
    \includegraphics{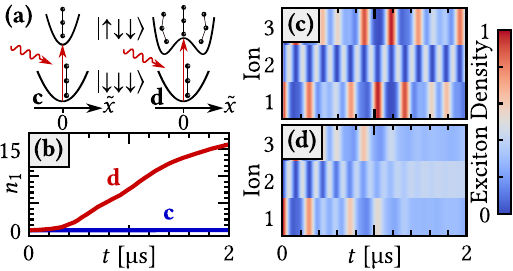}
    \caption{
        \textbf{Exciton dynamics in the presence of a conformational change.} (a) Sketch of the molecular potentials associated with the states $\ket*{\ed\ed\ed}$ and $\ket*{\eu\ed\ed}$, as a function of the reaction coordinate $\tilde{x}$ defined in Fig.~\ref{fig:MainPicture}(e). The red arrows indicate the electronic excitation of ion 1 from $\ket*{\ed}$ to $\ket*{\eu}$ induced by an incoming photon. The left sketch shows the scenario where the equilibrium geometry after the excitation is unchanged. Conversely, in the scenario displayed on the right, the equilibrium position, i.e., the potential minimum, is shifted. The corresponding spatially resolved exciton transport dynamics is shown in panels (c) and (d), respectively. (b) Time evolution of the phonon occupation number $n_1$ in the mode $q_1$ for the two scenarios. (c) Time evolution of the exciton densities $\rho_k$ governed by Hamiltonian \eqref{eq:ThreeIonModel} for the scenario where the equilibrium position is not shifted. (d) Analogous dynamics to (c), but for the scenario where the equilibrium position is shifted. For the simulation, $n=50$ (principal quantum number of $^{88}\mathrm{Sr}^+$), $\alpha=10^9\,\mathrm{Vm^{-2}}$, $\beta=0.25\times10^7\,\mathrm{Vm^{-2}}$, and $\Delta^{(2)}=\Delta^{(3)}=0$, leading to $J=-4.5$, $\varepsilon_\eu=-1.4$, $\varepsilon_\ed=4.6$. For (c) $\nuRF=32\,\mathrm{MHz}$ and for (d) $\nuRF=64\,\mathrm{MHz}$.
        }
    \label{fig:Relaxation}
\end{figure}
}
\def\figTransitions{
\begin{figure}[t]
    \centering
    \includegraphics{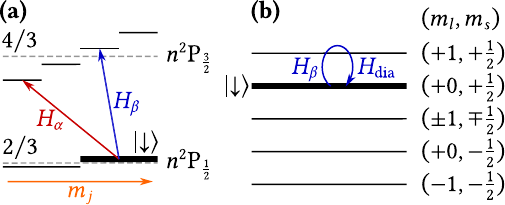}
    \caption{
        \textbf{Level structure of Rydberg P manifold.} (a) Relevant quadrupole transitions in the Rydberg P manifold in the weak-field Zeeman regime. The energy level associated with the state $\ket*{\ed}$ is indicated by a thicker line. Here, $m_j$ is the total magnetic quantum number, which labels the Zeeman sublevels. The red arrow shows the transition driven by $H_\alpha$, Eq.~\eqref{eq:AlphaHamiltonian}, and the blue arrow the transition driven by $H_\beta$, Eq.~\eqref{eq:BetaHamiltonian}. The fractions denote the Land\'e factors of the levels and the Zeeman splitting is drawn to scale. (b) Analogous to panel (a), but in the Paschen-Back regime. The energy levels are associated with electronic states in the uncoupled basis. The states are labelled by the magnetic quantum numbers $m_l$ and $m_s$, associated with the orbital and spin angular momentum. In this regime, the diamagnetic term $H_\mathrm{dia}$ and the contribution $H_\beta$, Eq.~\eqref{eq:BetaHamiltonian}, lead to energy shifts.
    }
    \label{fig:Transitions}
\end{figure}
}
\def\figElectronicParameters{
\begin{figure}[t]
    \centering
    \includegraphics{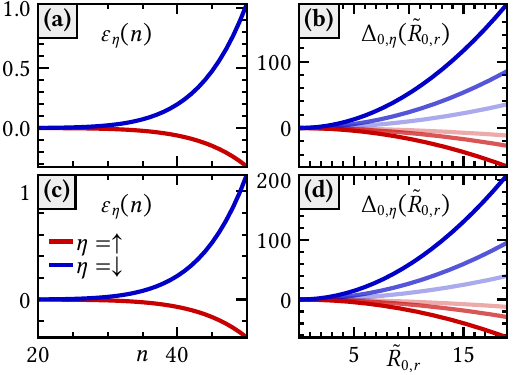}
    \caption{\textbf{Electronic parameters.} (a) Numerically computed $\varepsilon_\eta$, defined in Eq.~\eqref{eq:VarepsilonEta}, as a function of the principal quantum number $n$ for \srp \ \cite{code}. For the electric field gradients we used the parameter values $\alpha=10^9\,\mathrm{Vm^{-2}}$ and $\beta=10^7\,\mathrm{Vm^{-2}}$. The magnetic field was set to $B_0=20\,\mathrm{mT}$. Finally, the series in Eq.~\eqref{eq:VarepsilonEta} was truncated (including boundaries) at $n\pm7$. (b) Detunings given in Eq.~\eqref{eq:GeometricDetunings} resulting from the crystal's geometry, as a function of the radial ion distance from the trap axis ($z$-axis) $\tilde{R}_{0,r}=R_{0,r}\slash\xhoz$ at equilibrium. We plot the detunings for $n=40,45,50$, where more opaque lines correspond to larger principal quantum numbers. The parameters $\alpha$, $\beta$, and $B_0$ were the same as in panel (a). (c) Analogous to panel (a), but with $B_0=250\,\mathrm{mT}$, such that the system is approximately in the Paschen-Back regime. To numerically compute the radial matrix elements in the Paschen-Back regime, we use the fine-structure states with the lowest total angular momentum quantum number $j$, e.g., for the transition between $\ket*{\eu}$ and $\ket*{\ed}$ we use $\mud=\mel*{n^2\mathrm{S}_\frac{1}{2}}{r}{n^2\mathrm{P}_\frac{1}{2}}$ \cite{code}. This is justified because, for \srp, $n=50$, and $\mud'=\mel*{n^2\mathrm{S}_\frac{1}{2}}{r}{n^2\mathrm{P}_\frac{3}{2}}$, we find $(\abs*{\mud}-\abs*{\mud'})\slash\abs*{\mud}\approx1.4\,\%$. (d) Analogous to panel (b), but with the magnetic field used in panel (c).
    }
    \label{fig:ElectronicParameters}
\end{figure}
}
\def\figRelevanceOfQuarticTerm{
\begin{figure}[t]
    \centering
    \includegraphics{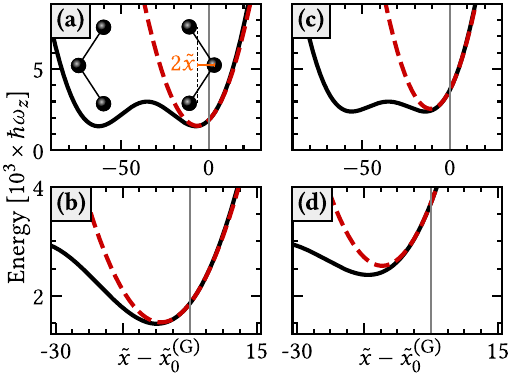}
    \caption{\textbf{Molecular potential energy surfaces.} (a) Potential energy (black solid line) given by Eq.~\eqref{eq:StateDependentPotential} for the special case $N=3$, as a function of the reaction coordinate $\tilde{x}$. The potential is evaluated along the curve $\gamma(\tilde{x})$ (see text). Here, $\tilde{x}^\mathrm{(G)}_0$ is the ground-state equilibrium value of the reaction coordinate $\tilde{x}$, the trap anisotropy was set to $a^2_x=2$, and all ions are in the Rydberg P state $\ket*{\ed}$ with polarisability $\varepsilon_\ed=0.5$. Red dashed parabolas show the harmonically approximated molecular potential in the P manifold, defined in Eq.~\eqref{eq:ModelSIUnits}. This approximation captures the full potential (black solid line) best, close to the point $\tilde{x}=\tilde{x}^{(\mathrm{G})}_0$ (grey lines). (b) Zoom into panel (a). (c) Analogous plot to panel (a), but with $\varepsilon_\ed=1$. (d) Zoom into panel (c).
    }
    \label{fig:RelevanceOfQuarticTerm}
\end{figure}
}
\def\figFacilitation{
\begin{figure}[t]
    \centering
    \includegraphics{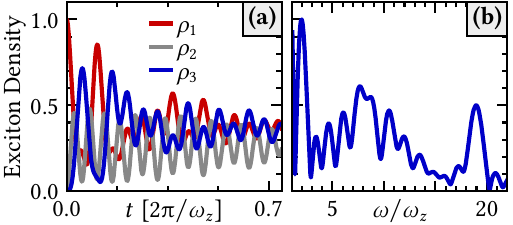}
    \caption{\textbf{Exciton transport facilitated by molecular vibrations.} (a) Time evolution of exciton densities $\rho_k$ at ion $k$ for an initial state of the form $\ket*{\eu\ed\ed}\otimes\ket*{\alpha_\mathrm{cs},0,0}$. Here, the exciton is localised on ion 1 and the collective mode $q_1$ is in the coherent-state $\ket*{\alpha_\mathrm{cs}}$ with amplitude $\alpha_\mathrm{cs}=7.7$, approximately equal to the amplitude used for panel (b) in Fig.~\ref{fig:Blockage} from the main text. In contrast to Fig.~\ref{fig:Blockage} from the main text, here we choose $\Delta^{(2)}$ to be the negative potential barrier introduced by the molecular vibrations (see text). (b) Fourier spectrum of the exciton density at the third ion, $\rho_3(t)$. The peak on the right of the spectrum is more pronounced compared to the one in Fig.~\ref{fig:Blockage}(c) in the main text. This is because the detuning $\Delta^{(2)}$ cancels the potential barrier introduced by the vibrations and thereby facilitates the exciton transport.
    }
    \label{fig:Facilitation}
\end{figure}
}
%%%%%%%%%%%%%%%%%%%%%%%%%%%%%%%%%%%%%%%%%%%%%%%%%%%%%%%%%%%%%%%%%%%%%%%%%%%%%%%%

%%% TITLE
\title{Non-equilibrium exciton dynamics in tailored molecular potentials of Rydberg ion crystals}

%%% AUTHOR INFORMATION
\author{Simon Euchner}
\author{Mathias B. M. Svendsen}
\affiliation{Institut f\"ur Theoretische Physik and Center for Integrated Quantum Science and Technology, Universit\"at Tübingen, Auf der Morgenstelle 14, 72076 T\"ubingen, Germany}
\author{Igor Lesanovsky}
\affiliation{Institut f\"ur Theoretische Physik and Center for Integrated Quantum Science and Technology, Universit\"at Tübingen, Auf der Morgenstelle 14, 72076 T\"ubingen, Germany}
\affiliation{School of Physics and Astronomy and Centre for the Mathematics and Theoretical Physics of Quantum Non-Equilibrium Systems, The University of Nottingham, Nottingham, NG7 2RD, United Kingdom}

%%% ABSTRACT
\begin{abstract}
    Trapped ions excited to high-lying electronic states combine strongly coupled collective vibrational and electronic degrees of freedom with long-ranged interparticle interactions. These ingredients enable the quantum simulation of biochemical processes, associated with the dynamics of excitons in non-perturbative parameter regimes. The key feature of such a quantum simulator are electronic-state-dependent molecular potential surfaces which can be strongly coupled. This allows to shed light on a variety of mechanisms underlying exciton transport. We illustrate this in a system of three trapped ions, which is amenable to an \emph{ab initio} treatment. Given that ion traps can be routinely prepared with hundreds of ions, these quantum simulators can immediately realise scenarios which are inaccessible by current numerical methods.
\end{abstract}

\maketitle

%%% SECTION

\textit{Introduction.} Understanding and exploiting the transport of excitation energy across molecules \cite{foerster1959} is of significant scientific and technological relevance \cite{cao2020,engel2007,lambert2013,bardeen2014}. Concrete examples are organic photovoltaics \cite{scholes2010} and \emph{excitonic wires} \cite{zhou2022}, a tool to guide excitation energy, so-called \emph{excitons}, over long distances. Shedding light on their dynamics poses challenges due to the complex interplay of molecular vibrations and electronic degrees of freedom. Generally, a full analysis of this \emph{vibronic} coupling necessitates the consideration of exponentially large state spaces, making \emph{ab initio} calculations intractable \cite{atkins2011}. This is especially the case in regimes where perturbative treatments, such as adiabatic approximations \cite{born1927,pope1999} or an effective description via master equations \cite{eisfeld2012}, fail.

Analogue quantum simulation \cite{manin1980,feynman1982} provides access to these regimes. Initial efforts explore highly excited atoms, so-called Rydberg atoms, to study micrometer-sized diatomic molecules \cite{boisseau2002,overstreet2009,bendkowsky2009,kiffner2012,sassmanhausen2016,shaffer2018,exner2025}, molecular instabilities \cite{euchner2025}, and the Jahn-Teller effect \cite{magoni2023,jahn1937,jahn1938}. Further, they have been used to study excitation energy transport \cite{muelken2007,guenter2013,schempp2014,yang2019,kosior2023,raupach2026}, the formation of Rydberg aggregates and clusters \cite{wuester2018,schempp2014}, probed with the so-called \emph{van der Waals explosion} \cite{faoro2016}, as well as the Su-Schrieffer-Heeger model \cite{hague2012}. On the other hand, collective vibrations, native to trapped ion setups, have been employed to study electron transport dynamics \cite{so2024} and vibrationally assisted exciton transfer \cite{padilla2025,ramm2014,gorman2018}, using phonon-mediated electronic interactions \cite{sorensen1999,sorensen2000}. Trapped Rydberg ions \cite{mueller2008,schmidtkaler2011} combine collective vibrational degrees of freedom with the exaggerated properties of Rydberg states \cite{higgins2019,higgins2019B}. Hitherto, Rydberg ions have been used to study electronic-state-dependent conformational changes \cite{li2012,mallweger2025}, conical intersections \cite{gambetta2021,whitlow2023,valahu2023,belfakir2026}, and spectral signatures of vibronic coupling \cite{wilkinson2024}. Importantly, all these phenomena have in common that they are linked to physical processes that take place in strongly-coupled electronic state-dependent \emph{molecular potential surfaces} --- the electronic state-dependent potentials experienced by the nuclei of a molecule \cite{atkins2011}.

In this work, we illustrate how trapped Rydberg ions allow to shed light on the physics of exciton transport processes that rely on coupled molecular potential surfaces. We first present the class of model Hamiltonians that are implementable on this quantum simulation platform. Subsequently we focus on a system consisting of three ions. Here, we showcase several non-equilibrium processes, among them motional-state-dependent excitonic transport, vibrationally assisted exciton transfer \cite{gorman2018}, and excitonic relaxation induced by a molecular conformational change. These examples highlight that quantum simulation with Rydberg ion crystals may indeed serve to gather important insights into fundamental physical mechanisms, which are of cross-disciplinary relevance, for instance energy transport across the \ac{fmo} protein \cite{fenna1975,olson2004,nalbach2011}.

%%% SECTION

%%% FIGURE
\figMainPicture

\textit{Quantum simulator architecture.}  We consider a chain of $N$ alkaline-earth metal ions of mass $M$ confined in the harmonic potential of a linear Paul trap with trap frequencies $\omega_x$, $\omega_y$, and $\omega_z$ in the respective spatial directions; see Refs.~\cite{paul1990,leibfried2003,schmidtkaler2011} and Fig.~\ref{fig:MainPicture}(a). Each ion is described by a two-level system with states $\ket*{\eu}$ and $\ket*{\ed}$ encoded in a Rydberg S and P state, respectively, as shown in Fig.~\ref{fig:MainPicture}(b). Due to the interplay between the trap confinement and the Coulomb repulsion, so-called \emph{Coulomb crystals} form at sufficiently low temperature \cite{morigi2025}. Here, the motion of the ions is described by small displacements of the collective modes of the crystal. To keep the presentation concise, we start by presenting the Hamiltonian that is realised in this quantum simulator setup; its detailed derivation is provided in the \ac{sm} \cite{supmat}:
\begin{multline}\label{eq:Model}
    H
    =
    \frac{1}{2}
    \big[
        \bm{p}^2+\bm{q}^\mathrm{T}\Gamma^2\bm{q}
    \big]
    +
    \sum^N_{u=1}\sum_{\eta=\eu,\ed}
    \big[
        \delta^{(u)}_\eta
        +
        \varepsilon_\eta
        W^{(u)}(\bm{q})
    \big]
    P^{(u)}_\eta
\\
    +
    \sum^N_{u=1}\sum^N_{v(\neq{u})=1}
    J^{(u,v)}
    \big[S^{(u)}_+S^{(v)}_-+\mathrm{h.\,c.}\big]
\,.
\end{multline}
The first term describes the ion crystal's $3N$ collective vibrational modes $\bm{q}$ and their conjugate momenta $\bm{p}$, where $\Gamma=\mathrm{diag}(\Gamma_1,\dots,\Gamma_{3N})$ is a diagonal matrix that contains the associated eigenfrequencies. We measure $\bm{q}$ in units of the harmonic oscillator length $\xhoz=\sqrt{\hbar\slash(M\omega_z)}$, frequencies in units $\omega_z$, and we set $\hbar\equiv1$.

In the context of this work we interpret the first term of Hamiltonian (\ref{eq:Model}) as the motion of nuclei within a (harmonic) molecular potential surface \cite{atkins2011}. The second term contains contributions to this molecular potential which are electronic state dependent, reflected by the projectors $P^{(u)}_\eta=\dyad*{\eta}$. These entail constant energy shifts, represented by the detunings $\delta^{(u)}_\eta$, and modifications of the potential shape given by
\begin{equation}\label{eq:StateDependentPotentialsMainText}
    W^{(u)}(\bm{q})
    =
    \frac{1}{2}\bm{q}^\mathrm{T}A^{(u)}\bm{q}
    -
    \bm{f}^{(u)}\cdot\bm{q}
\,.
\end{equation}
The overall strength of these modifications is controlled by the ions' electric polarisability $\varepsilon_\eta$, which depends on the considered Rydberg state $\ket*{\eta}$. As shown in Refs.~\cite{higgins2019,pokorny2020,mallweger2025}, the polarisability can be controlled experimentally, e.g., by applying external electromagnetic fields. The matrix $A^{(u)}$ in Eq.~\eqref{eq:StateDependentPotentialsMainText} affects the frequency and the structure of the vibrational modes. The vector $\bm{f}^{(u)}$ represents mechanical forces, which emerge because the equilibrium geometry, which is determined by a stationary point of the state-dependent molecular potential, depends on the electronic state. Therefore, after electronic excitation, the ion crystal is not necessarily at equilibrium and hence experiences a force that displaces its collective motional modes. The final term in Hamiltonian \eqref{eq:Model} describes excitation transport, which is driven by resonant dipole-dipole (`flip-flop') interactions at coherent rate $J^{(u,v)}$. To model this process, we employ exciton creation and annihilation operators: $S_+=\dyad*{\eu}{\ed}$ and $S_-=\dyad*{\ed}{\eu}$, respectively.

The state-dependent energy shifts $\delta^{(u)}_\eta$ (detunings), the modifications to the molecular potentials $W^{(u)}(\bm{q})$ in Eq.~\eqref{eq:StateDependentPotentialsMainText}, and the transport rates $J^{(u,v)}$ are all determined by the ion crystal's geometry. In turn, the geometry is controlled by the parameters of the Paul trap, i.e. the radio frequency $\nuRF$ and the electric field gradients $\alpha$ and $\beta$ \cite{mueller2008,schmidtkaler2011}. When showing numerical data we will provide these parameter values, which are throughout chosen such that they correspond to those of current experiments conducted with \srp ions \cite{higgins2019}. In general, the coupling constants in Hamiltonian \eqref{eq:Model} are determined by the specific experimental setup and in particular the ionic species, as we discuss in the \ac{sm} \cite{supmat}. However, in order to be self-contained, we outline in the \ac{em} how each coupling constant in Hamiltonian \eqref{eq:Model} can be computed.

%%% SECTION

\textit{Simplified setting.} For large ion crystals the model Hamiltonian \eqref{eq:Model} is numerically intractable. To nevertheless obtain an understanding concerning the type of physical processes that are captured by it, we consider a simplified setting, which consists of three ions, arranged in a linear chain, and three relevant vibrational modes. A sketch of these modes is shown in Fig.~\ref{fig:MainPicture}(d). Note that we solely consider transversal collective vibrational modes, because, for typical parameters, the coupling to the longitudinal modes is much weaker in a linear Paul trap, as detailed in the \ac{sm} \cite{supmat}. With regard to the electronic degrees of freedom, we restrict ourselves to the single-excitation subspace shown in Fig.~\ref{fig:MainPicture}(c). This subspace is spanned by the three basis states $\ket*{1}=\ket*{\eu\ed\ed}$, $\ket*{2}=\ket*{\ed\eu\ed}$, and $\ket*{3}=\ket*{\ed\ed\eu}$, which represent an exciton localized on ions 1, 2, and 3, respectively. Within these approximations, Hamiltonian \eqref{eq:Model} reads (see \ac{sm} \cite{supmat})
\begin{eqnarray}\label{eq:ThreeIonModel}
    H   & = &
    \sum^3_{k=1} \bigg[\frac{\bm{p}^2}{2}+V^{(k)}(\bm{q})\bigg] \dyad*{k}
    +
    J\sum^3_{k>l} \frac{\dyad*{k}{l}+\dyad*{l}{k}}{\abs*{k-l}^3} \nonumber
\\
    &&+
    \Delta^{(2)} \dyad*{2} +\Delta^{(3)} \dyad*{3}
\,.
\end{eqnarray}
The first term describes the motion of the ions in the molecular potentials associated with the states $\ket*{k}$,
\begin{equation}\label{eq:BosonicHamiltonians}
    V^{(k)}(\bm{q})
    =
    \frac{1}{2}\bm{q}^\mathrm{T}\Gamma^2\bm{q}
    +
    \varepsilon_\eu W^{(k)}(\bm{q})
    +
    \varepsilon_\ed \sum_{l\neq{k}} W^{(l)}(\bm{q})
\,.
\end{equation}
The second term of Eq.~\eqref{eq:ThreeIonModel} describes exciton transport, which is parametrised by the coherent transport rate $J$. The detunings $\Delta^{(2)}$ and $\Delta^{(3)}$ represent constant energy shifts of the molecular potential surfaces. Both  $\Delta^{(2)}$ and $\Delta^{(3)}$ can be controlled via Stark shifts and weak inhomogeneous magnetic fields.

%%% SECTION

%%% FIGURE
\figBlockage

\textit{Scenarios of exciton transport.} To illustrate the capabilities of the quantum simulator we study three relevant scenarios, described by Eq.~\eqref{eq:ThreeIonModel}. Throughout, we quantify the dynamics via the exciton densities $\rho_k(t)=\abs*{\braket*{k}{\psi(t)}}^2$ ($\ket*{\psi(t)}$ is the time-evolved state of the system), which measure the population at ion $k$.

The first scenario is a simple model for a light-harvesting linear molecule, in which an exciton is created by an incident photon, as depicted in Fig.~\ref{fig:MainPicture}(c). When transferring momentum to the molecule, the photon `kicks' a collective mode, here $q_1$, such that it undergoes oscillations described by the motional coherent state $\ket*{\alpha_\mathrm{cs}}$. Interestingly, as shown in Fig.~\ref{fig:Blockage}, the time an exciton needs to move through the chain increases with the momentum transferred by the photon, i.e., with increasing coherent-state amplitude $\alpha_\mathrm{cs}$.

Remarkably, the motional-state dependence of the transport speed allows a simple analytical description: in the limit of weakly coupled vibrational and electronic dynamics, the initial motional wave-packet is approximately unperturbed by the exciton dynamics and the vibrations follow a classical trajectory. Under these assumptions one obtains the effective Hamiltonian
\begin{equation}\label{eq:MinimalModel}
    H_\mathrm{eff}
    =
    J(\dyad{1}{2}+\dyad*{2}{3}+\mathrm{h.\,c.})
    +
    \frac{\alpha^2_\mathrm{cs}}{4}
    \frac{\varepsilon_\eu-\varepsilon_\ed}{\sqrt{\Gamma^2_1+\varepsilon_\ed}}
    \dyad*{2}
\,,
\end{equation}
with details being provided in the \ac{sm} \cite{supmat}. This effective model directly shows that the oscillations manifest in an effective potential barrier (energy shift on ion 2) through which an exciton has to tunnel on its way from ion 1 to ion 3. The height of this potential barrier increases with the `kick' amplitude $\alpha_\mathrm{cs}$. Consequently, the tunnelling rate, and with it the speed of exciton transport, decreases.

%%% GRAUE LINIE UND DELTA_MAX SYMBOL

%%% FIGURE
\figVAET

However, molecular vibrations can also facilitate exciton transport. Such \emph{vibrationally assisted exciton transfer} can occur when the dipolar `flip-flop' processes strongly couple to the molecular vibrations \cite{gorman2018}. To illustrate this, we consider the scenario described in Fig.~\ref{fig:VAET}(a). Here, we use the energy offset $\Delta$ to control the energetic separation between the molecular potentials associated with different positions of the exciton in the ion crystal. The vibrationally assisted transport is demonstrated in Fig.~\ref{fig:VAET}(b), which shows the time-averaged exciton density at ion 3 (the exciton is initially created at ion 1) and the time-averaged phonon occupation number. Both curves acquire a maximum at the same energy offset $\Delta_\mathrm{max}$, which is a signature of vibrationally assisted transport. Spatially resolved data is shown in Fig.~\ref{fig:VAET}(c,d) for two different values of the energy offset, where (c) shows the vibrationally assisted regime. Indeed, exciton transport takes place when $\Delta=\Delta_\mathrm{max}$, while the exciton remains localised at its initial position when $\Delta$ becomes too large. Note that for too small energy offsets $\Delta$, exciton transport takes place but is not vibrationally assisted. This is visible in Fig.~\ref{fig:VAET}(b), which shows that in this regime indeed ion 3 gets excited, but the vibrational mode $q_1$ does not get populated.

%%% FIGURE
\figRelaxation

A further important mechanism of molecular dynamics that impacts exciton transport is a state-dependent change of the molecular conformation, in our case, from a linear to a bent chain. In the context of trapped ions, such a conformational change can take place when the equilibrium geometry of the ion crystal depends on the electronic state of individual ions. Recently, this scenario has been reported in Refs.~\cite{li2012,mallweger2025}. Here, we consider a linear ion crystal in which an incoming photon excites ion 1 from state $\ket*{\ed}$ to state $\ket*{\eu}$. This amounts to the creation of an exciton localised on ion 1, analogous to the excitation process shown in Fig.~\ref{fig:MainPicture}(c). We distinguish two scenarios which are depicted in Fig.~\ref{fig:Relaxation}(a): one scenario where the equilibrium position is not shifted upon exciton creation, and one where it shifts from a linear to a zig-zag chain. In both cases the conformation of the ion crystal is unchanged immediately after the exciton is created. This is a consequence of the Franck-Condon principle \cite{franck1926,condon1928,atkins2011}, which states that during electronic excitation the molecular geometry remains unchanged. In the case when the equilibrium position is not shifted, the ion crystal after excitation is at equilibrium and the exciton transport is only weakly affected by the molecular vibrations. This is indicated by the numerical simulation of the phonon mode occupation in the mode $q_1$ shown in Fig.~\ref{fig:Relaxation}(b), which stays approximately zero over time. In this case, the spatially resolved transport shown in Fig.~\ref{fig:Relaxation}(c) shows that the exciton coherently moves back and forth between ions 1 and 3. When the equilibrium geometry is shifted once the exciton is created, the ion crystal begins to vibrate, which leads to an increase in the number of phonons. This is shown in Fig.~\ref{fig:Relaxation}(b) exemplarily for one mode. This introduces motional decoherence, which ultimately stops the coherent (oscillating) exciton transport. The density distribution of the exciton across the ions then becomes approximately stationary, as shown in Fig.~\ref{fig:Relaxation}(d). 

%%% SECTION

\textit{Summary and outlook.} Trapped Rydberg ions allow to shed light on the physics of exciton transport processes that rely on coupled molecular potential surfaces. Ultimately, such a quantum simulator may be used to treat technologically important systems of cross-disciplinary relevance, such as organic photovoltaics \cite{scholes2010}, excitonic wires \cite{zhou2022}, and light-harvesting complexes such as the \ac{fmo} protein \cite{fenna1975,olson2004}. This is due to the variety of molecular shapes that can be realised by exciting selected ions from a larger two-dimensional ion crystal, which can be confined in a Penning trap \cite{vogel2024,martins2026}. Finite-temperature effects in experiments \cite{schmidtkaler2002,eschner2003} further allow to study temperature-dependent transport rates \cite{banerjee2016}, as well as controlled molecular dynamics out of thermal equilibrium. Additionally, the modifications to the molecular potential caused by Rydberg excitation give rise to geometrical distortions of the ion crystal. Such distortions can support the formation of excitonic polarons --- a bound state between an exciton and the crystal distortions \cite{holstein1959,silbey1980,dai2024}. This shows that Rydberg ions offer a promising route to probe their fundamental properties in large molecular complexes.

%%% SECTION

\textit{Acknowledgements.} We thank J. W. P. Wilkinson and W. S. Martins for fruitful discussions on Rydberg ions. We thank M. Moroder for suggesting relevant work on the FMO complex. We acknowledge support from the QuantERA II programme (project CoQuaDis, DFG Grant No. 532763411) that has received funding from the EU H2020 research and innovation programme under GA No. 101017733. We also acknowledge funding from the Deutsche Forschungsgemeinschaft (DFG, German Research Foundation) through the Research Unit FOR 5413/1, Grant No.~465199066, from the Leverhulme Trust (Grant No. RPG-2024-112), from the Centre for Integrated Quantum Science and Technology (IQST Boosting Programme) and from the European Union through the ERC grant OPEN-2QS (Grant No. 101164443).

%%% SECTION

\textit{Data availability.} The data and code that support the findings of this work are openly available \cite{code}.

%%% SECTION

\textit{AI usage.} The authors used \emph{Claude (Sonnet 4.6)} for grammar and spell-checking of the manuscript. All suggestions were reviewed and approved by the authors.

%%% BIBLIOGRAPHY
\bibliography{references}

%%% END MATTER
\section{End Matter}

\textit{Coupling constants.} Here, we outline how the coupling constants in the Hamiltonians \eqref{eq:Model} and \eqref{eq:ThreeIonModel} can be derived from the Paul trap radio frequency $\nuRF$ and electric field gradients $\alpha$, $\beta$. These parameters are well explained in Refs.~\cite{paul1990,leibfried2003,schmidtkaler2011}. However, in order to be self-contained, we briefly outline their meaning here: a linear Paul trap consists of a combination of a quadrupole static and a quadrupole oscillating electric field. This field configuration leads to harmonic confinement in all three spatial directions, with the associated trap frequencies $\omega_x$, $\omega_y$, and $\omega_z$. In terms of $\nuRF$, $\alpha$, and $\beta$ these frequencies are
\begin{equation}
    \omega_{x,y}
    =
    \sqrt{2\bigg[\frac{e \alpha}{2 \pii \nuRF M}\bigg]^2
          -
          \frac{\omega^2_z}{2}}
\,, \ \ \
    \omega_z
    =
    \sqrt{\frac{4e\beta}{M}}
\,.
\end{equation}
The oscillating component of the electric field is responsible for confinement in the radial directions $x$ and $y$. It oscillates at the radio frequency $\nuRF{\sim}10\,\mathrm{MHz}$ and its strength is determined by the electric field gradient $\alpha{\sim}10^9\,\mathrm{V\slash{m^2}}$. The static electric field leads to confinement along the $z$-direction. Its magnitude is determined by the electric field gradient $\beta{\sim}10^7\,\mathrm{V\slash{m^2}}$. For instance, typical parameter values for $\nuRF$, $\alpha$, and $\beta$, like the ones stated here, can be found in Ref.~\cite{schmidtkaler2011}.

The first coupling constants we consider are the collective mode frequencies $\Gamma$. These are determined by the molecular potential experienced by the centre-of-mass positions of the ions. Explicitly, for an ion crystal with all ions in their ground state, this potential reads
\begin{multline}\label{eq:ExternalPotentialEndMatter}
    V(\bm{R})
    =
    \frac{1}{2}M\omega^2_z
    \sum^N_{u=1}\sum_{i=x,y,z}
    a^2_i \big(R^{(u)}_i\big)^2
\\
    +
    \frac{e^2}{8\pii\varepsilon_0}
    \sum^N_{u=1}\sum^{N}_{v(\neq{u})=1}
    \frac{1}{\norm*{\bm{R}^{(u)}-\bm{R}^{(v)}}}
\,.
\end{multline}
The first contribution describes the harmonic confinement introduced by the trap. Here, $a_i=\omega_i\slash\omega_z$ ($i=x,y,z$) are the trap anisotropies in all three spatial directions. The centre-of-mass position of ion $u$ is described by the position vector $\bm{R}^{(u)}=[R^{(u)}_x,R^{(u)}_y,R^{(u)}_z]$. For convenience, we collect all of the position vectors in the $3N$-dimensional vector $\bm{R}=[\bm{R}^{(1)},\dots,\bm{R}^{(N)}]$. The second term in Eq.~\eqref{eq:ExternalPotentialEndMatter} describes the Coulomb repulsion between the ions of charge $e>0$, where $\varepsilon_0$ is the vacuum permittivity. From the potential in Eq.~\eqref{eq:ExternalPotentialEndMatter} we can numerically calculate the equilibrium ion positions $\bm{R}_0$ via the equilibrium condition
\begin{equation}\label{eq:EquilibriumConditionEndMatter}
    \pdv{V}{R_\mu}\bigg\vert_{\bm{R}_0} = 0
\,,
\end{equation}
for all $\mu=1,\dots,3N$. Here, $R_\mu$ denotes component $\mu$ of the $3N$-dimensional vector $\bm{R}$. For the cold ion crystals we consider in this work, the displacements of the ions from their equilibrium positions are small compared to inter-ion distances. This allows to expand the potential in Eq.~\eqref{eq:ExternalPotentialEndMatter} in the small displacements $\delta\bm{R}=\bm{R}-\bm{R}_0$. This approximation amounts to
\begin{equation}\label{eq:ApproximatedPotentialEndMatter}
    V(\bm{R})
    \approx
    V(\bm{R}_0)
    +
    \frac{1}{2} M \omega^2_z
    \delta\bm{R}^\mathrm{T}
    K
    \delta\bm{R}
\,,
\end{equation}
where $K$ is the Hessian matrix of the potential in Eq.~\eqref{eq:ExternalPotentialEndMatter} at equilibrium (in units of $M\omega^2_z$) with components
\begin{equation}\label{eq:HessianEndMatter}
    K_{\mu,\nu}
    =
    \pdv[2]{V}{R_\mu}{R_\nu}\bigg\vert_{\bm{R}_0}
\,.
\end{equation}
In the main text, we express the potential \eqref{eq:ApproximatedPotentialEndMatter} in the collective eigenmodes $\bm{q}$ of the ion crystal. To obtain these eigenmodes we diagonalise the Hessian matrix with a rotation $O$:
\begin{equation}\label{eq:RotationEndMatter}
    \Gamma^2
    =
    O^\mathrm{T}
    K
    O
\,.
\end{equation}
Here, the diagonal matrix $\Gamma$ contains the eigenfrequencies of the collective modes, which, using the rotation $O$, are defined as
\begin{equation}
    \bm{q} = \frac{1}{\xhoz} O^\mathrm{T} \delta\bm{R}
\,.
\end{equation}
Except for special cases, the matrices $O$ and $\Gamma$ must be computed numerically. Note that for ion crystals at equilibrium, all eigenfrequencies are positive, because the equilibrium $\bm{R}_0$ is a minimum. Therefore, the diagonal matrix $\Gamma=\sqrt{\Gamma^2}$ is well-defined.

Next, we focus on the coupling constants $\bm{f}^{(u)}$ and $A^{(u)}$ in Eq.~\eqref{eq:StateDependentPotentialsMainText} from the main text. These can be calculated with the rotation $O$ defined in Eq.~\eqref{eq:RotationEndMatter}. Explicitly, we find the expressions
\begin{equation}\label{eq:FormulaForForcesEndMatter}
    f^{(u)}_\mu
    =
    - \frac{1}{\xhoz} R^{(u)}_{0,x} O_{(u,x),\mu}
    - \frac{1}{\xhoz} R^{(u)}_{0,y} O_{(u,y),\mu}
\end{equation}
and
\begin{equation}\label{eq:FormulaForCurvaturesEndMatter}
    A^{(u)}_{\mu,\nu}
    =
    O^\mathrm{T}_{\mu,(u,x)}O_{(u,x),\nu}
    +
    O^\mathrm{T}_{\mu,(u,y)}O_{(u,y),\nu}
\,.
\end{equation}
Here, we introduced the shorthand notations $(u,x)=3(u-1)+1$ and $(u,y)=3(u-1)+2$ ($u=1,\dots,N$) to label the components of the rotation $O$ which correspond to the spatial direction $i$ of ion $u$. Further, $R^{(u)}_{0,i}$ ($i=x,y$) are the equilibrium positions of ion $u$ in the two radial directions. A full derivation of expressions \eqref{eq:FormulaForForcesEndMatter} and \eqref{eq:FormulaForCurvaturesEndMatter} is given in the \ac{sm} \cite{supmat}.

The next coupling constants we consider are the transport rates $J^{(u,v)}$ in Hamiltonian \eqref{eq:Model} from the main text. These result from the dipole-dipole interaction between the two Rydberg ions $u$ and $v$ and can be computed with the formula
\begin{equation}\label{eq:HoppingRateEndMatter}
    \hbar\omega_z
    J^{(u,v)}
    =
    \frac{e^2}{8\pii\varepsilon_0}
    \frac{\bm{\mu}^\dagger_\mathrm{d}
          S^{(u,v)}
          \bm{\mu}_\mathrm{d}}
         {\norm*{\bm{R}^{(u)}_0-\bm{R}^{(v)}_0}^3}
\,.
\end{equation}
In this formula, $\bm{\mu}_\mathrm{d}=\mel*{\eu}{\bm{r}}{\ed}$ is the dipole matrix element between the Rydberg S and P states $\ket*{\eu}$ and $\ket*{\ed}$, and $\bm{R}^{(u)}_0$ is the position vector of ion $u$ at equilibrium. The matrix $S^{(u,v)}$ can be calculated via the formula
\begin{equation}
    S^{(u,v)}_{i,j}
    =
    1
    -
    2
    \frac{\big(R^{(u)}_{0,i}-R^{(v)}_{0,i}\big)\big(R^{(u)}_{0,j}-R^{(v)}_{0,j}\big)}
         {\norm*{\bm{R}^{(u)}_0-\bm{R}^{(v)}_0}^2}
\,.
\end{equation}
Finally, the transport rate $J$ in Hamiltonian \eqref{eq:ThreeIonModel} in the main text is defined as $J=J^{(1,2)}=J^{(2,3)}$.

Next, we focus on the coupling constants $\varepsilon_\eu$ and $\varepsilon_\ed$, which are proportional to the polarisabilities of the Rydberg states $\ket*{\eu}$ and $\ket*{\ed}$. In the units employed in this work, we obtain the expression
\begin{equation}
    \varepsilon_\eta
    =
    \sum_{\eta'\neq\eta}
    \frac{\alpha}{\beta}
    \frac{e\alpha\abs*{\mel*{\eta}{r_x}{\eta'}}^2}{E_\eta-E_{\eta'}}
\,.
\end{equation}
This result follows directly from second-order perturbation theory and the sum runs over the full single-ion spectrum consisting of the states $\ket*{\eta'}$ with energies $E_{\eta'}$. The energies and transition matrix elements are computed numerically. To evaluate the sum, we truncated the series after numerical convergence was reached.

Finally, we consider the detunings $\delta^{(u)}_\eta$ in Hamiltonian \eqref{eq:Model} from the main text. These are given by the formula
\begin{equation}\label{eq:DetuningsEndMatter}
    \delta^{(u)}_\eta
    =
    \Delta^{(u)}_{0,\eta}
    +
    \Delta^{(u)}_\eta
\,.
\end{equation}
The first term is a detuning which is controlled by the geometry of the ion crystal. In the regime where $\abs*{R^{(u)}_{0,x}},\abs*{R^{(u)}_{0,y}}$ are on the order of a few to a few tens of harmonic oscillator lengths, $\xhoz$, these detunings can be computed with the formula
\begin{equation}
    \Delta^{(u)}_{0,\eta}
    =
    \frac{\varepsilon_\eta}{2\xhoz^2}
    \big[ \big(R^{(u)}_{0,x}\big)^2
          +
          \big(R^{(u)}_{0,y}\big)^2 \big]
\,,
\end{equation}
which we derive in the \ac{sm} \cite{supmat}. Note that for linear ion crystals $\Delta^{(u)}_{0,\eta}=0$. The second term in Eq.~\eqref{eq:DetuningsEndMatter} can be controlled via Stark shifts and inhomogeneous magnetic fields. In terms of these detunings, $\Delta^{(2)}$ and $\Delta^{(3)}$ in Eq.~\eqref{eq:ThreeIonModel} from the main text read:
\begin{equation}
    \Delta^{(k)}
    \equiv
    \Delta^{(k)}_\eu - \Delta^{(k)}_\ed
    -
    \Delta^{(1)}_\eu + \Delta^{(1)}_\ed
\end{equation}
with $k=2,3$.

%%% SUPPLEMENTAL MATERIAL
\clearpage
\setcounter{secnumdepth}{3}
\setcounter{section}{0}
\setcounter{equation}{0}
\setcounter{figure}{0}
\setcounter{table}{0}
\renewcommand{\thesection}{S\arabic{section}}
\renewcommand{\theequation}{S\arabic{equation}}
\renewcommand{\thefigure}{S\arabic{figure}}
\renewcommand{\thetable}{S\arabic{table}}
\onecolumngrid
\begin{center}
        \large\bf
        Supplemental material for:
\\
        Non-equilibrium exciton dynamics in tailored molecular potentials of Rydberg ion crystals
\end{center}
\vspace*{0mm}
\begin{center}
    \begin{minipage}{15cm}
        \begin{center}
            Simon Euchner$^1$, Mathias B. M. Svendsen$^1$, and Igor Lesanovsky$^{1,2}$
\vspace*{2mm}
\\
            $^1$\textit{Institut f\"ur Theoretische Physik and Center for Integrated Quantum Science and Technology,
\\
    Universit\"at T\"ubingen, Auf der Morgenstelle 14, 72076 T\"ubingen, Germany}
\\
            $^2$\textit{School of Physics and Astronomy and Centre for the Mathematics
\\
            and Theoretical Physics of Quantum Non-Equilibrium Systems,
\\
            The University of Nottingham, Nottingham, NG7 2RD, United Kingdom}
        \end{center}
    \end{minipage}
\end{center}

\vspace*{0mm}
\begin{center}
    \begin{minipage}{14.5cm}
        \hspace*{1em} Here we present details on results from the main text. In the first Section we give a detailed derivation of Eq.~\eqref{eq:Model} from the main text and discuss the result. In the second Section we derive Eq.~\eqref{eq:ThreeIonModel} from the main text and give analytical formulae for the collective-mode frequencies and collective modes of an ion crystal consisting of three ions. In the third Section we derive Eq.~\eqref{eq:MinimalModel} from the main text. In the fourth Section, we analytically solve the Schr\"odinger equation for Hamiltonian \eqref{eq:MinimalModel}, which allows to quantify the transport speed's dependence on the potential barrier.
    \end{minipage}
\end{center}
\vspace*{5mm}
\twocolumngrid

%%% SECTION
\section{Derivation of the simulator Hamiltonian}
\label{app:DerivationOfTheSimulatorHamiltonian}

To derive the Hamiltonian \eqref{eq:Model} from the main text we start with a singly-ionised alkaline-earth-metal atom (Group 2 \cite{pse}). Such an ion can be approximately described as a two-body problem, composed of the valence electron and an \emph{effective nucleus}. This effective nucleus consists of the nucleus itself and $Z-2$ electrons, which are strongly bound to the nucleus ($Z$ is the core charge). The dynamics of this effective nucleus and the valence electron can be described with the Hamiltonian
\begin{equation}
    H_\mathrm{ion}
    =
    \frac{\pen^2}{2\men}
    +
    \frac{\pe^2}{2\me}
    +
    V_\mathrm{eff}(\norm*{\re-\ren})
\,.
\end{equation}
The first term describes the kinetic energy of the effective nucleus (`en') with charge $+2e>0$. The second term is the electron's (`e') kinetic energy. The third term is an effective potential which accounts for the finite size and polarisability of the effective nucleus, as described in Ref.~\cite{aymar1996}. We further include spin-orbit coupling in this effective potential. To be concrete, throughout this work we use the form of the effective potential employed in Ref.~\cite{wilkinson2025}. Using this effective potential, we solve the time-independent Schr\"odinger equation for the energy spectra and eigenstates of the ions. For this task we employ the software \emph{AlkCalc} \cite{alkcalc,code}.

Next, we assume that the ion is confined in a linear Paul trap \cite{paul1990,leibfried2003,schmidtkaler2011}. Close to the centre of such a trap, an ion at position $\bm{\xi}=[\xi_x,\xi_y,\xi_z]^\mathrm{T}$ experiences the time-dependent electric quadrupole potential \cite{mueller2008,schmidtkaler2011,li2012},
\begin{multline}\label{eq:ElectricPotential}
    \Phi(\bm{\xi},t)
    =
    \alpha \cos(\omgRF t) (\xi^2_x-\xi^2_y)
\\
    -
    \beta ([1+\epsilon]\xi^2_x+[1-\epsilon]\xi^2_y-2\xi^2_z)
\,.
\end{multline}
Here, $\beta{\sim}10^7\,\mathrm{Vm^{-2}}$ and $\alpha{\sim}10^9\,\mathrm{Vm^{-2}}$ are the electric field gradients of the static and \ac{rf} electric field, respectively, where the latter oscillates at frequency $\nu_\mathrm{RF}=\omgRF\slash(2\pii){\sim}\text{10-100}\,\mathrm{MHz}$ \cite{mueller2008,schmidtkaler2011}. Further, $0\leq\epsilon<1$ is an anisotropy parameter which can be controlled with the linear Paul trap's geometry. In addition to the electric fields of the trap, we introduce a static homogeneous magnetic field $\bm{B}=B_0\bm{e}_z$, aligned with the trap and quantisation axis. To introduce this field, we employ the standard minimal coupling substitution: $\pe\mapsto\pe+e\bm{A}$, $\pen\mapsto\pen-2e\bm{A}$, with the vector potential $\bm{A}(\bm{r})=(\bm{B}\times\bm{r})\slash{2}$. Therefore, the ion subject to the external magnetic fields and the external electric fields of the trap, can be described by the Hamiltonian
\begin{multline}
    H(t)
    =
    \frac{(\pen-2e\bm{A}(\ren))^2}{2\men}
    +
    \frac{(\pe+e\bm{A}(\re))^2}{2\me}
\\
    +
    2e\Phi(\ren,t)
    -
    e\Phi(\re,t)
\\
    V_\mathrm{eff}(\norm*{\re-\ren})
    -
    \muB \elgf B_0 s_z
\,,
\end{multline}
where $\muB$ is Bohr's magneton, $\elgf\approx2$ is the electron's gyromagnetic ratio, and $s_z$ is the electron's spin projection operator onto the quantisation axis. In the following it will be convenient to describe the ion using its centre-of-mass coordinates $\bm{R}=[R_x,R_y,R_z]^\mathrm{T}$ and its relative coordinates $\bm{r}=\re-\ren=[r_x,r_y,r_z]^\mathrm{T}$. Further, we perform the gauge-transformation \cite{schmelcher1991}
\begin{equation}
    G
    =
    \ee^{-\ii f(\bm{R},\bm{r})}
\,, \
    f(\bm{R},\bm{r}) = \frac{e(\men+2\me)}{M} \bm{A}(\bm{R}) \cdot \bm{r}
\,,
\end{equation}
where $M=\men+\me$ denotes the ion's total mass. Finally, analogously to Refs.~\cite{cook1985,mueller2008,wilkinson2025}, we transform into a rotating frame via the unitary operator
\begin{equation}
    U(t)
    =
    \ee^{-\ii g(\bm{R},t)}
\,, \
    g(\bm{R},t) = \frac{e\alpha\sin(\omgRF t)}{\hbar\omgRF} \big[R^2_x-R^2_y\big]
\,.
\end{equation}
This yields a transformed Hamiltonian (for simplicity again called $H$) that decomposes into four parts:
\begin{align}
    H(t)
    &\mapsto
    U^\dagger(t) G^\dagger H(t) G U(t) - \ii\hbar U^\dagger(t) \dot{U}(t)
\\
    &\approx \label{eq:SingleIonHamiltonian}
    H_\mathrm{ext}(t)
    +
    H_\mathrm{int}
    +
    H_\mathrm{ext\text{-}int}(t)
    +
    H_\mathrm{mm}(t)
\,,
\end{align}
where the approximation sign signals that we ignored contributions of order $\me\slash{M}{\sim}10^{-5}$.

The first term in Eq.~\eqref{eq:SingleIonHamiltonian} describes the external motion of the ion's centre-of-mass in a harmonic trapping potential, and is given by the expression
\begin{equation}\label{eq:ExternalHamiltonian}
    H_\mathrm{ext}
    =
    \frac{\bm{P}^2}{2M}
    +
    \frac{1}{2} M \omega^2_z \sum_{i=x,y,z} a^2_i R^2_i
    -
    \frac{\omgc}{2} L_z
\,.
\end{equation}
Here, $L_z$ is the centre-of-mass angular momentum in the axial direction and the frequency anisotropies, defined as $a_i=\omega_i\slash\omega_z$ ($i=x,y,z$), are
\begin{equation}\label{eq:TrapAnisotropies}
    a_{x,y}
    =
    \sqrt{\bigg[\frac{\sqrt{2}e\alpha}{\omega_z \omgRF M}\bigg]^2
          -
          \frac{1\pm\epsilon}{2}
          +
          \bigg[\frac{\omgc}{2\omega_z}\bigg]^2}
\,, \ \
    a_z = 1
\,,
\end{equation}
where we chose the axial trapping frequency $\omega_z=\sqrt{4e\beta\slash{M}}$ as the frequency scale. The negative term appearing in the expressions of the $x$- and $y$-anisotropies $a_{x,y}$ is a manifestation of the fact that the static electric field anti-traps the ion in the radial direction. The cyclotron frequency $\omgc=eB_0\slash{M}$ enhances the radial trapping frequency, analogously to a Penning trap \cite{vogel2024}.

The second term in Eq.~\eqref{eq:SingleIonHamiltonian} describes the internal dynamics of the valence electron, and is given by the expression
\begin{multline}\label{eq:InternalHamiltonian}
    H_\mathrm{int}(t)
    =
    \frac{\bm{p}^2}{2m}
    +
    V_\mathrm{eff}(r)
    -
    e \Phi(\bm{r},t)
\\
    +
    \muB B_0 (l_z+\elgf s_z)
    +
    \frac{e^2 B^2_0}{8m} (r^2_x+r^2_y)
\,.
\end{multline}
Here, $\Phi$ is the electric quadrupole potential \eqref{eq:ElectricPotential} and $l_z$ is the orbital angular momentum of the valence electron.

The third term in Eq.~\eqref{eq:SingleIonHamiltonian} couples the centre-of-mass external motion to the internal dynamics of the valence electron:
\begin{multline}\label{eq:CouplingHamiltonian}
    H_\mathrm{ext\text{-}int}(t)
    =
    2 e \alpha [g_x(t) r_x R_x + g_y(t) r_y R_y] - 4e\beta r_z R_z
\\
    +
    e \alpha \sin(\omgRF t) \frac{\omgc}{\omgRF} (r_x R_y + r_y R_x)
\\
    +
    \omgc (r_x P_y - r_y P_x)
\,.
\end{multline}
Here, we defined the dimensionless functions
\begin{equation}\label{eq:gFunctions}
    g_{x(y)}(t)
    =
    \mp\cos(\omgRF{t})+\beta(1\pm\epsilon)\slash\alpha-M\omgc^2\slash{(4e\alpha)}
\,,
\end{equation}
which account for the time-dependence of the electric field that oscillates at \ac{rf} $\nuRF$.

The final term in Eq.~\eqref{eq:SingleIonHamiltonian} describes so-called \emph{micromotion}, i.e., fast oscillatory motion of the ions resulting from the \ac{rf} electric field. For details on micromotion in Paul traps, please see Refs.~\cite{cook1985,mueller2008,martins2025,wilkinson2025}. Explicitly, for $H_\mathrm{mm}$ in Eq.~\eqref{eq:SingleIonHamiltonian} we obtain
\begin{multline}
    H_\mathrm{mm}
    =
    -
    \frac{e^2\alpha^2}{M\omgRF^2} \cos(2\omgRF t) \big[R^2_x+R^2_y\big]
\\
    -
    \frac{2e\alpha}{M\omgRF} \sin(\omgRF t) [R_x P_x - R_y P_y]
\\
    -
    2 e \alpha \sin(\omgRF t) \frac{\omgc}{\omgRF} R_x R_y
\,.
\end{multline}
At sufficiently fast \ac{rf} $\nuRF$ and ions close to the trap's axis (here, the $z$-axis), the amplitudes of the oscillations are small and the contribution can be omitted, as is shown in Refs.~\cite{cook1985,martins2025}.

In the following we focus on moderate magnetic fields with $\abs*{B_0}{\sim}0\text{--}300\,\mathrm{mT}$. For the singly-ionised strontium isotope \srp this leads to cyclotron frequencies $\omgc\approx2\pii\times0\text{-}0.05\,\mathrm{MHz}$. We will later see that the typical simulation times we are interested in are on the order of 1\,\textmu s. Even for $\abs*{B_0}=300\,\mathrm{mT}$ the angle by which the crystal rotates is less than $20^\circ$. We note that for heavier ions, such as \bap, the rotation of the crystal is less pronounced, as $\omgc{\sim}M^{-1}$. Therefore, we can ignore the rotation for small magnetic fields $B_0$ and heavy ions, which amounts to neglecting the final term in Eq.~\eqref{eq:ExternalHamiltonian}. The final term in Eq.~\eqref{eq:CouplingHamiltonian} has the same structure as $L_z$ and can therefore be omitted as well. Finally, we note that for our magnetic fields (maximally) $\omgc\slash\omgRF{\sim}10^{-2}$. Because of this, we can ignore the penultimate contribution in Eq.~\eqref{eq:CouplingHamiltonian}, as it is at least two orders of magnitude weaker than the term proportional to $e\alpha{g}_i(t)$. As a result, we arrive at the single-ion Hamiltonian
\begin{align}
    H(t)
    &=
    \frac{\bm{P}^2}{2M}
    +
    \frac{1}{2} M \omega^2_z \sum_{i=x,y,z} a^2_i R^2_i \nonumber
    +
    \frac{\bm{p}^2}{2m}
    +
    V_\mathrm{eff}(r) \nonumber
\\
    &-
    e \Phi(\bm{r},t)
    +
    \muB B_0 (l_z+\elgf s_z)
    +
    \frac{e^2 B^2_0}{8m} (r^2_x+r^2_y) \nonumber
\\
    &+
    2 e \alpha [g_x(t) r_x R_x + g_y(t) r_y R_y] - 4e\beta r_z R_z
\,.
\end{align}

Next, we consider $N$ ions loaded into the same linear Paul trap. This scenario is described by the Hamiltonian
\begin{equation}
    H(t)
    =
    \sum^N_{u=1} H^{(u)}(t)
    +
    H_\mathrm{mb}
\,.
\end{equation}
Here, $H^{(u)}(t)$ denotes the single-ion Hamiltonian in Eq.~\eqref{eq:SingleIonHamiltonian} associated with the $u$-th ion. The long-ranged Coulomb repulsion between the ions is described by the many-body Hamiltonian
\begin{multline}
    H_\mathrm{mb}
    =
    \frac{1}{4\pii\varepsilon_0}
    \sum^N_{u=1}\sum^{N}_{v(\neq{u})=1}
    \frac{1}{2}
    \bigg[
        \frac{(2e)(2e)}
             {\norm*{\bm{r}^{(u)}_\mathrm{en}-\bm{r}^{(v)}_\mathrm{en}}}
\\
        +
        \frac{(-e)(2e)}
             {\norm*{\bm{r}^{(u)}_\mathrm{e}-\bm{r}^{(v)}_\mathrm{en}}}
        +
        \frac{(2e)(-e)}
             {\norm*{\bm{r}^{(u)}_\mathrm{en}-\bm{r}^{(v)}_\mathrm{e}}}
        +
        \frac{(-e)(-e)}
             {\norm*{\bm{r}^{(u)}_\mathrm{e}-\bm{r}^{(v)}_\mathrm{e}}}
    \bigg]
\,.
\end{multline}
In this work, we approximate the problem as electrostatic. This is valid for non-relativistic ions, where effects such as Bremsstrahlung and retardation are not appreciable. Moreover, typical ion distances are on the order of a few micrometres, whereas the size of Rydberg ions scales as $\aB{n}^2$ \cite{higgins2019}, with the Bohr radius $\aB\approx 53\,\mathrm{pm}$, yielding only ${\sim}200\,\mathrm{nm}$ even for principal quantum numbers as large as $n=60$. This justifies a multipole expansion of $H_\mathrm{mb}$. To second order we find
\begin{equation}\label{eq:ManyBodyHamiltonian}
    H_\mathrm{mb}
    \approx
    H_\mathrm{cc}
    +
    H_\mathrm{cd}
    +
    H_\mathrm{dd}
    +
    H_\mathrm{cq}
\,.
\end{equation}
The terms, in the order presented above, are the Coulomb repulsion between the ions,
\begin{equation}\label{eq:ChargeChargeInteraction}
    H_\mathrm{cc}
    =
    \ecz
    \sum^N_{u=1}
    \sum^{N}_{v(\neq{u})=1}
    \frac{1}{\norm*{\bar{\bm{R}}^{(u,v)}}}
\,,
\end{equation}
the charge-dipole interaction
\begin{equation}\label{eq:ChargeDipoleInteraction}
    H_\mathrm{cd}
    =
    \ecz \vepse
    \sum^N_{u=1}
    \sum^N_{v(\neq{u})=1}
    \frac{2 \bar{\bm{R}}^{(u,v)}}
         {\norm*{\bar{\bm{R}}^{(u,v)}}^3}
    \cdot
    \frac{\bm{r}^{(u)}}{\aB}
\,,
\end{equation}
the dipole-dipole interaction
\begin{equation}\label{eq:DipoleDipoleInteraction}
    H_\mathrm{dd}
    =
    \ecz \vepse^2
    \sum^N_{u=1}
    \sum^N_{v(\neq{u})=1}
    \frac{\bm{r}^{(u)}}{\aB}
    \cdot
    \frac{S(\bar{\bm{R}}^{(u,v)})}
         {\norm*{\bar{\bm{R}}^{(u,v)}}^3}
    \frac{\bm{r}^{(v)}}{\aB}
\,,
\end{equation}
and, finally, the interaction between one ion's charge and the quadrupole of another one
\begin{equation}
    H_\mathrm{cq}\label{eq:ChargeQuadrupoleInteraction}
    =
    \ecz \vepse^2
    \sum^N_{u=1}\sum^N_{v(\neq{u})=1}
    \frac{\bm{r}^{(u)}}{\aB}
    \frac{S(\bar{\bm{R}}^{(u,v)})}
         {\norm*{\bar{\bm{R}}^{(u,v)}}^3}
    \cdot
    \frac{\bm{r}^{(u)}}{\aB}
\,.
\end{equation}
In the above expressions we introduced the three-dimensional matrix (for some vector $\bm{\xi}$)
\begin{equation}\label{eq:StructureMatrix}
    S(\bm{\xi})
    =
    1_3 - 3 \frac{\bm{\xi}\otimes\bm{\xi}^\mathrm{T}}{\norm*{\bm{\xi}}^2}
\,.
\end{equation}
Further, we introduced the distance vectors $\bm{R}^{(u,v)}=\bm{R}^{(u)}-\bm{R}^{(v)}$ and the characteristic energy scale of an ion crystal oriented along the axial direction, $\ecz=M\omega^2_z\ell^2_z\slash{2}$, where $\ell_z=[e^2\slash(4\pii\varepsilon_0M\omega^2_z)]^{1\slash{3}}$ is the length scale of typical inter-ion distances. Finally, `bar'-ed quantities are scaled by a factor of $\ell_z$, i.e., $\bar{\xi}=\xi\slash\ell_z$ (for some quantity $\xi$), and we introduced the parameter $\vepse=\aB\slash\ell_z$.

Combining Eqs.~\eqref{eq:ExternalHamiltonian}, \eqref{eq:InternalHamiltonian}, \eqref{eq:CouplingHamiltonian} with the many-body interaction in Eq.~\eqref{eq:ManyBodyHamiltonian} yields the full Hamiltonian for the trapped ion
crystal,
\begin{equation}\label{eq:FullHamiltonian}
    H(t)
    =
    H_\mathrm{ext}
    +
    H_\mathrm{int}(t)
    +
    H_\mathrm{int\text{-}ext}(t)
\,,
\end{equation}
again decomposed into an external (vibrational), an internal (electronic), and a coupling (vibronic) term. To compare the magnitudes of different contributions in the following discussion, we introduce, in addition to $\vepse$, the parameter $\vepsv=\xhoz\slash\ell_z$, with the harmonic oscillator length $\xhoz=\sqrt{\hbar\slash(M\omega_z)}$. In the following we only keep terms of second order, i.e., we omit $\vepsv^3$, $\vepsv^2\vepse$, $\vepsv\vepse^2$, $\vepse^3$, and higher-order corrections. We justify this by the following observation: each order $\vepse$ comes with an electronic matrix element between Rydberg states, i.e., roughly a factor of $10^3$ (see scaling of dipole and quadrupole transition matrix elements with $n$ in \cite{higgins2019}), such that for typical values $\ell_z{\sim}1\,\text{\textmu{m}}$ we have $\ev*{\norm*{\bm{r}}}\vepse{\sim}0.05$. On the other hand, $\xhoz\slash\ell_z{\sim}0.01$, for a typical value $\xhoz{\sim}10\,\text{nm}$. Therefore, $\vepse{\sim}\vepsv$, such that third-order terms like $\vepse^2\vepsv$ can be discarded consistently. In the following, we expand each part of Hamiltonian \eqref{eq:FullHamiltonian} separately.

%%% SUBSECTION
\subsection{External Hamiltonian}
\label{subsec:ExternalHamiltonian}

The external Hamiltonian is obtained by combining the external single-ion contributions in Eq.~\eqref{eq:ExternalHamiltonian} with the charge-charge interaction in Eq.~\eqref{eq:ChargeChargeInteraction}:
\begin{equation}\label{eq:ManyBodyExternalHamiltonian}
    H_\mathrm{ext}
    =
    \sum^N_{u=1} H^{(u)}_\mathrm{ext} + H_\mathrm{cc}
\,.
\end{equation}
At zero temperature, the interplay of the harmonic confinement introduced by the linear Paul trap and the long-ranged Coulomb repulsion between the ions leads to a static crystal structure, a so-called \emph{Coulomb crystal} \cite{morigi2025}. For sufficiently cold ion crystals, this justifies an expansion in small displacements around the equilibrium ion positions, which are a solution to the equation
\begin{equation}\label{eq:EquilibriumCondition}
    \pdv{V_\mathrm{ext}}{\bar{\bm{R}}}\bigg\vert_{\bar{\bm{R}}_0}
    =
    \bm{0}
\,,
\end{equation}
with the potential experienced by the external degrees of freedom
\begin{equation}\label{eq:GroundStateExternalfPotential}
    \frac{V_\mathrm{ext}(\bm{R})}{\ecz}
    =
    \sum^N_{u=1}\sum_{i=x,y,z}
    a^2_i \big(\bar{R}^{(u)}_i\big)^2
    +
    \sum^N_{u=1}\sum^{N}_{v(\neq{u})=1}
    \frac{1}{\norm*{\bar{\bm{R}}^{(u,v)}}}
\,.
\end{equation}
Here, we introduced the $3N$-dimensional position vector $\bm{R}^\mathrm{T}=[(\bm{R}^{(1)})^\mathrm{T},\dots,(\bm{R}^{(N)})^\mathrm{T}]$ to keep the notation compact. Close to the equilibrium position $\bm{R}_0$, we may approximate the many-body external potential $V_\mathrm{ext}$ in the small displacements $\delta\bm{R}=\bm{R}-\bm{R}_0$. To second order, we find
\begin{equation}\label{eq:ApproximatedPotential}
    V_\mathrm{ext}(\bm{R})
    \approx
    V_\mathrm{ext}(\bm{R}_0)
    +
    \ecz \delta\bar{\bm{R}}^\mathrm{T} K \delta\bar{\bm{R}}
\,,
\end{equation}
where the Hessian matrix of the potential $V_\mathrm{ext}$ in units of $M\omega^2_z$ is
\begin{equation}
    K
    =
    \frac{1}{2}
    (\nabla^\mathrm{T}_{\bar{\bm{R}}} \otimes \nabla_{\bar{\bm{R}}})
    \vert_{\bm{R}_0}
    \frac{V_\mathrm{ext}}{\ecz}
\,.
\end{equation}
Note that $\norm*{\delta\bar{\bm{R}}}\slash(3N){\sim}\vepsv$, such that Eq.~\eqref{eq:ApproximatedPotential} includes contributions up to (including) order $\ecz\vepsv^2$. Next, we transform to the collective vibrational modes $\bar{\bm{Q}}=O^\mathrm{T}\delta\bar{\bm{R}}$, where $O$ is a rotation that diagonalises $K$, i.e., $O^\mathrm{T}KO=\Gamma^2$, and $\hbar\omega_z\Gamma=\hbar\omega_z\mathrm{diag}(\Gamma_1,\dots,\Gamma_{3N})$ is the diagonal matrix containing the collective mode eigenfrequencies. After applying the transformation to the collective vibrational modes, the potential \eqref{eq:GroundStateExternalfPotential} assumes the form [after subtracting the constant energy $V(\bm{R}_0)$]
\begin{equation}\label{eq:FinalExternalHamiltonian}
    H_\mathrm{ext}
    \approx
    \frac{\bm{P}^2}{2M}
    +
    \ecz \bar{\bm{Q}}^\mathrm{T} \Gamma^2 \bar{\bm{Q}}
\,.
\end{equation}
Note that each component $\bar{Q}_\mu$ ($\mu=1,\dots,3N$) of $\bar{\bm{Q}}$ is of order $\vepsv$ and higher-order contributions are omitted.

%%% SUBSECTION
\subsection{Internal Hamiltonian}
\label{subsec:InternalHamiltonian}

The next step is to expand the internal Hamiltonian in Eq.~\eqref{eq:FullHamiltonian} to second order in $\vepse$ and $\vepsv$. This contribution is composed of the single-body internal terms in Eq.~\eqref{eq:InternalHamiltonian}, and the many-body dipole-dipole, Eq.~\eqref{eq:DipoleDipoleInteraction}, and charge-quadrupole, Eq.~\eqref{eq:ChargeQuadrupoleInteraction}, interactions:
\begin{equation}\label{eq:ManyBodyInternalHamiltonian}
    H_\mathrm{int}
    =
    \sum^N_{u=1}H^{(u)}_\mathrm{int}(t)
    +
    H_\mathrm{dd}
    +
    H_\mathrm{cq}
\,.
\end{equation}
In this work, we consider two regimes: one regime (i) where the effects due to the magnetic field $B_0$ are much weaker than the spin-orbit interaction (weak-field regime), and one regime (ii) where the contributions due to the magnetic field dominate (strong-field/Paschen-Back regime).

We first consider the weak-field regime (i). In this regime, the diamagnetic term (the one proportional to $B^2_0$) in Eq.~\eqref{eq:InternalHamiltonian} can be ignored and the spin-orbit term dominates over the Zeeman splitting between energy levels with different magnetic quantum number $m_j$. Therefore, the Hamiltonian describing the $N$ ions in the magnetic field without the external electric fields of the Paul trap takes the form
\begin{equation}\label{eq:DiagonalIonHamiltonian}
    H_\mathrm{ions}
    =
    \sum_{\bm{\eta}} E_{\bm{\eta}}  \dyad*{\bm{\eta}}
\,,
\end{equation}
with the collective eigenstates $\ket*{\bm{\eta}}=\ket*{\eta^{(1)},\dots,\eta^{(N)}}$. The quantum numbers $\eta^{(u)}=(n,l,j,m_j)$ label the states in the fine-structure basis. These states are fixed by the principal quantum number $n$, the orbital angular momentum quantum number $l$, the total angular momentum quantum number $j$, and the magnetic quantum number $m_j$. The single-ion eigenenergy associated with the quantum number $\eta$ is $E_{\eta}=E_{n,l,s,j}+\lande(l,s,j)\muB{B}_0m_j$, where $g_\mathrm{L}(l,s,j)$ is the Land\'e factor. In this work, we reduce Hamiltonian \eqref{eq:DiagonalIonHamiltonian} to a two-level system, encoded in the two Rydberg states $\ket*{\eu}\equiv\ket*{n^2\mathrm{S}_\frac{1}{2}(\frac{1}{2})}$ and $\ket*{\ed}\equiv\ket*{n^2\mathrm{P}_\frac{1}{2}(\frac{1}{2})}$.

Next, we consider the Paschen-Back regime (ii). This regime is reached for sufficiently high $n$ and sufficiently large $B_0$. Here, the spin-orbit coupling is negligible compared to the Zeeman shifts (this is a particularly good approximation for light ions and large magnetic fields). In this case the ions' Hamiltonian \eqref{eq:DiagonalIonHamiltonian} is approximately diagonal in the \emph{uncoupled basis}, that is, the basis where $l$ and $s$ are not coupled. Therefore, the quantum number $\eta=(n,l,m_l,s,m_s)$ labels the electronic states, with the magnetic quantum numbers $m_l$ and $m_s$. Analogously to the weak-field regime (i), we describe each ion as a two-level system. However, here we choose the states $\ket*{\eu}\equiv\ket*{n^2\mathrm{S}(m_l=0,m_s=\frac{1}{2})}$ and $\ket*{\ed}\equiv\ket*{n^2\mathrm{P}(m_l=0,m_s=\frac{1}{2})}$. We remark that the crossover between the weak-field-Zeeman (i) and the Paschen-Back (ii) regime for a Rydberg ion is neatly depicted in Ref.~\cite{martins2026}, which shows the electronic spectrum as a function of the magnetic field $B_0$.

In both regimes (i) and (ii), the electronic Hilbert space associated with the $N$ ions splits into well-isolated resonant subspaces with energies $E_\nexc=\nexc(E_\mathrm{\eu}-E_\mathrm{\ed})$, where $\nexc$ is the number of ions in the electronic state $\ket*{\eu}$, i.e., the number of excitons. The resonant subspaces are visualised in Fig.~\ref{fig:MainPicture}(c) in the main text, where we show the electronic state space for the case $N=3$.

With the electronic states $\ket*{\eu}$ and $\ket*{\ed}$ fixed in both the weak-field (i) and in the Paschen-Back (ii) regime, we proceed to simplify the internal Hamiltonian in Eq.~\eqref{eq:ManyBodyInternalHamiltonian}. Apart from the diagonal contribution in Eq.~\eqref{eq:DiagonalIonHamiltonian} this Hamiltonian also contains the interaction of the ions with the electric fields of the Paul trap, the dipole-dipole interaction, and the interaction of the quadrupole moment of one ion with the charge of another one. Generally, these interactions can couple the states in Eq.~\eqref{eq:DiagonalIonHamiltonian} and the ions cannot be described as simple two-level systems. This is the case when the energy splitting between subspaces of different $\nexc$ is not much larger than the magnitude of the additional terms. Because of this, the goal of this Section is twofold: one goal is to expand each term to second order in $\vepse$ and $\vepsv$, and the other goal is to determine when $\nexc$ remains a good quantum number. To this end, we consider each contribution in Eq.~\eqref{eq:ManyBodyInternalHamiltonian} separately.

The first contribution we consider is the dipole-dipole interaction in Eq.~\eqref{eq:DipoleDipoleInteraction}. Regardless of the regimes (i) and (ii), our choice of states leads to the expression
\begin{equation}\label{eq:dipoleDipoleInteractionInBasis}
    \frac{H_\mathrm{dd}}{\hbar\omega_z}
    \approx
    \sum^N_{u=1}
    \sum^N_{v(\neq{u})=1}
    J^{(u,v)}
    \big[ S^{(u)}_+ S^{(v)}_-
          +
          S^{(u)}_- S^{(v)}_+ \big]
\,,
\end{equation}
with the local exciton creation and annihilation operators $S^{(u)}_+=\dyad*{\eu}{\ed}^{(u)}$, $S^{(u)}_-=\dyad*{\ed}{\eu}^{(u)}$, the dipole-dipole coupling strengths
\begin{equation}\label{eq:FullHoppingRates}
    \hbar\omega_z
    J^{(u,v)}
    =
    2e\beta
    \frac{\bm{\mu}^\dagger_\mathrm{d}
          S(\bar{\bm{R}}^{(u,v)}_0)
          \bm{\mu}_\mathrm{d}}
         {\norm*{\bar{\bm{R}}^{(u,v)}_0}^3}
\,,
\end{equation}
and the dipole transition matrix element $\bm{\mu}_\mathrm{d}=\mel*{\eu}{\bm{r}}{\ed}$. We observe two things: the coupling strengths $J^{(u,v)}$ already are of second order in $\vepse$ and they only weakly mix subspaces of different $\nexc$. The latter observation is seen by the fact that even for weak magnetic fields $B_0=20\,\mathrm{mT}$ --- such that the system is in the weak-field regime (i) --- the Zeeman shifts are on the order of $\muB{B}_0\slash{h}\approx280\,\mathrm{MHz}$. However, the coupling strengths $J^{(u,v)}$ are typically much weaker \cite{mueller2008}. For instance, for the strontium isotope \srp and $n=50$ with $\beta=10^7\,\mathrm{Vm^{-2}}$ we obtain $J^{(u,u+1)}\slash{h}{\sim}14\,\mathrm{MHz}$, which is much smaller than $280\,\mathrm{MHz}$. Note that in the Paschen-Back regime (ii) the Zeeman shifts can be much larger than 280\,MHz such that the condition that $\nexc$ is a good quantum number is also fulfilled. Because of our choice of states, we can, independent of the regimes (i) and (ii), further simplify expression \eqref{eq:FullHoppingRates}: between the states $\ket*{\eu}$ and $\ket*{\ed}$ $\sigma_\pm$ transitions are dipole-forbidden and only $\pi$ transitions are allowed. As a result, the coupling strengths $J^{(u,v)}$ can be reduced to
\begin{equation}\label{eq:DipoleDipoleCouplingStrengthsBeforeExpansion}
    \hbar 
    \omega_z
    J^{(u,v)}
    =
    \frac{2}{3\sigma}\frac{e\beta\mu^2_\mathrm{d}}{\norm*{\bar{\bm{R}}^{(u,v)}_0}^3}
    \Bigg[
        1
        -
        3\frac{(\bar{R}^{(u,v)}_{0,z})^2}
              {\norm*{\bar{\bm{R}}^{(u,v)}_0}^2}
    \Bigg]
\,.
\end{equation}
In this expression we defined the (radial) dipole matrix element $\mud=\mel*{n^2\mathrm{S}_\frac{1}{2}}{r}{n^2\mathrm{P}_\frac{1}{2}}$ and the number $\sigma$, which is 3 in the weak-field regime (i) [originating from a squared Clebsch-Gordan coefficient] and 1 in the Paschen-Back regime (ii).

Second-order energy shifts leading to van der Waals-interactions are strongly suppressed by the large energy splitting between subspaces of different $\nexc$, compared to the dipole-dipole coupling strength $J^{(u,v)}$. However, due to their dramatic scaling with ${\sim}n^{11}$, they can become appreciable for high principal quantum numbers. However, in this work we omit the van der Waals interaction, because we are interested in more typical quantum numbers used in current experiments with \srp \ \cite{higgins2019}. Furthermore, we focus on ionic chains, i.e., linear ion crystals and ion crystals for which the radial equilibrium distance of each ion, measured with respect to the trap axis, is small compared to inter-ion distances. For typical ion crystals, this means that the equilibrium ion distance from the trap axis is on the order of a few harmonic oscillator lengths, i.e., $\abs*{R^{(u)}_{0,x}}{\sim}1\text{--}10\xhoz$. For further information on the different geometries of ion crystals and the structural transitions between them, please see Refs.~\cite{raizen1992,schiffer1993,fishman2008,block2000,li2012,mallweger2025}. The assumption $\abs*{R^{(u)}_{0,x}}{\sim}1\text{--}10\xhoz$ entails that corrections to the linear geometry (stemming from the `slightly' non-linear shape of the chain) are roughly on the order of $\vepsv$. Therefore, for the ion crystals considered in this work, i.e., ionic chains, an expansion to first order in $R^{(w)}_{0,x}\slash\ell_z$ increases the overall order by the factor $\vepsv$. In particular, for the dipole-dipole coupling strengths this results in  Eq.~\eqref{eq:DipoleDipoleCouplingStrengthsBeforeExpansion} being transformed into
\begin{equation}\label{eq:TransportRates}
    \hbar\omega_z
    J^{(u,v)}
    \approx
    -\frac{4}{3\sigma}
    \frac{e\beta\mu^2_\mathrm{d}}
         {\abs*{\bar{R}^{(u,v)}_{0,z}}^3}
    \sim
    \beta n^4
\,,
\end{equation}
scaling with the fourth power of the principal quantum number $n$. Typical magnitudes are $e\beta\mud\approx9\,\mathrm{MHz}$ (\srp, $n=45$, $\beta=10^7\,\mathrm{Vm^{-2}}$).

The remaining contributions in Eq.~\eqref{eq:ManyBodyInternalHamiltonian} are single-body [final term in Eq.~\eqref{eq:InternalHamiltonian}] and many-body [see Eq.~\eqref{eq:ChargeQuadrupoleInteraction}] quadrupole contributions. In the Paschen-Back regime we additionally have the diamagnetic term [final term in Eq.~\eqref{eq:InternalHamiltonian}] $H_\mathrm{dia}$. Therefore, we consider the perturbation
\begin{equation}\label{eq:QuadrupoleContributions}
    \sum^N_{u=1} \big[ -e \Phi(\bm{r}^{(u)},t) \big]
    +
    H_\mathrm{cq}
    +
    H_\mathrm{dia}
    \equiv
    H_\alpha
    +
    H_\beta
    +
    H_\mathrm{dia}
\,.
\end{equation}
Here, we introduced the Hamiltonian $H_{\alpha}$, which is proportional to the \ac{rf} electric field gradient $\alpha$, as well as $H_\beta$, which is proportional to $\beta$. Their explicit forms are
\begin{equation}\label{eq:AlphaHamiltonian}
    H_\alpha(t)
    =
    -e\alpha\cos(\omgRF t)
    \sum^N_{u=1}
    \big[\big(r^{(u)}_x\big)^2-\big(r^{(u)}_y\big)^2\big]
\end{equation}
and
\begin{multline}
    H_\beta
    =
    e\beta
    \sum^N_{u=1}
    \big([1+\epsilon]\big(r^{(u)}_x\big)^2+[1-\epsilon]\big(r^{(u)}_y\big)^2-2\big(r^{(u)}_z\big)^2\big]
\\
    +
    2e\beta
    \sum^N_{u=1}
    \bm{r}^{(u)}
    \cdot
    Q^{(u)}
    \bm{r}^{(u)}
\,,
\end{multline}
with the quadrupole tensor
\begin{equation}
    Q^{(u)}
    =
    \sum^N_{v(\neq{u})=1}
    \frac{
        S(\bar{\bm{R}}^{(u,v)}_0)
    }{
        \norm*{\bar{\bm{R}}^{(u,v)}_0}^3
    }
\,.
\end{equation}
Analogously to the dipole-dipole interaction in Eq.~\eqref{eq:DipoleDipoleInteraction}, all quadrupole terms are of order $\ecz\vepse^2$ already, such that for the ionic chains considered in this work, we find
\begin{equation}
    Q^{(u)}
    \approx
    \sum^N_{u(\neq{v})=1}
    \frac{1}{\abs*{\bar{R}^{(u,v)}_{0,z}}^3}
    \mqty[1 & & \\ & 1 & \\ & & -2]
\,,
\end{equation}
where higher-order terms are at least of order $\vepse^2\vepsv$. From this form of the quadrupole tensor it follows that $H_\beta$ simply is
\begin{multline}\label{eq:BetaHamiltonian}
    H_\beta
    \approx
    e\beta
    \sum^N_{u=1}
    \bigg(
        q^{(u)}
        \big[\big(r^{(u)}\big)^2-3\big(r^{(u)}_z\big)^2\big]
\\
        +
        \epsilon\big[\big(r^{(u)}_x\big)^2-\big(r^{(u)}_y\big)^2\big]
    \bigg)
\,,
\end{multline}
with the scalar quantity
\begin{equation}
    q^{(u)}
    =
    1
    +
    2
    \sum^N_{v(\neq{u})=1}
    \frac{1}{\abs*{\bar{R}^{(u,v)}_{0,z}}^3}
\,.
\end{equation}
We note that the term resulting from the trap's radial anisotropy, which is proportional to the structure constant $\epsilon$, has the same symmetry as Eq.~\eqref{eq:AlphaHamiltonian}. However, its magnitude is much smaller than $H_\alpha$, because $\beta\slash\alpha{\sim}10^{-2}$. Therefore we can ignore the anisotropy term in Eq.~\eqref{eq:BetaHamiltonian} and both $H_\alpha$ and $H_\beta$ have well-defined symmetries.

The next goal is to determine if under the influence of the quadrupole terms the total number of excitons, $\nexc$, is still a good quantum number.

%%% FIGURE
\figTransitions

We start with the Hamiltonian $H_\alpha$ in the weak-field regime (i). It couples the state $\ket*{\ed}$ to $\ket*{n^2\mathrm{P}_\frac{3}{2}(-\frac{3}{2})}$ \cite{mueller2008}, as depicted in Fig.~\ref{fig:Transitions}(a). The coupling matrix element has  magnitude $e\alpha\muq$, where $\muq=\mel*{n^2\mathrm{P}_\frac{1}{2}}{r^2}{n^2\mathrm{P}_\frac{3}{2}}$ is the radial quadrupole matrix element. For \srp at $n=50$, yielding $\muq\approx3.2\times10^{6}\aB^2$, and $\alpha=10^9\,\mathrm{Vm^{-2}}$ we find $e\alpha\muq\approx2.2\,\mathrm{GHz}$. On the other hand, for $B_0=20\,\mathrm{mT}$, the energy separation of the states is $E_{\eta}-E_\ed=\hbar\times2\pii\times3.1\,\mathrm{GHz}$, $\eta=(50,1,\frac{1}{2},\frac{3}{2},-\frac{3}{2})$, such that the transition is very much possible and $\nexc$ is not a good quantum number. Note that the strong coupling between these states was experimentally observed in \cite{higgins2021}. However, this could be solved in different ways: by reducing the principal quantum number (the energy splitting scales as $n^{-3}$ \cite{higgins2019}), or by employing heavier ions, e.g., the barium isotope \bap, or by increasing the strength of the magnetic field. For instance, with $B_0=250\,\mathrm{mT}$ and sufficiently high $n$, the system is in the Paschen-Back regime. Here, $H_\alpha$ does not drive transitions within the P manifold for our choice of $\ket*{\ed}$, because the symmetry dictates that this term only drives transitions with $\abs*{\Delta{m}_l}=2$. By definition, the shifts introduced by the magnetic field in the Paschen-Back regime are large compared to the spin-orbit term, such that it does not drive additional transitions. In conclusion, this means that under the influence of $H_\alpha$, the number of excitons, $\nexc$, remains a good quantum number, because all other transitions come with $\Delta{l}=2$, and thus are far detuned due to the large energetic splitting between states of different $l$. We remark that to make sure the Paschen-Back regime is reached, one can also employ lighter ions such as the calcium isotope \cap: for this species, the spin-orbit term is smaller compared to \srp.

Next, we consider $H_\beta$. For weak magnetic fields, $H_\beta$ couples $\ket*{\ed}$ to $\ket*{n^2\mathrm{P}_\frac{3}{2}(\frac{1}{2})}$, as is depicted in Fig.~\ref{fig:Transitions}(a). However, for $n=50$ and \bap the coupling strength is $e\beta\muq\slash{h}{\sim}20\,\mathrm{MHz}$ ($\beta=10^7\,\mathrm{Vm^{-2}}$), which is much smaller than the splitting of 9.3\,GHz between these states. In contrast, in the Paschen-Back regime, there are no transitions within the P manifold, as depicted in Fig.~\ref{fig:Transitions}(b). However, $\ket*{\ed}$ experiences a first-order energy shift with magnitude $e\beta\muq$. In conclusion, we find that $H_\beta$, like $H_\alpha$, does not alter the total number of excitons, $\nexc$.

The final contribution we consider is the diamagnetic term $H_\mathrm{dia}$. This term is only relevant in regime (ii), i.e., in the Paschen-Back regime. For magnetic fields in the range of a few Tesla and sufficiently large principal quantum numbers $n$, this term can even couple manifolds of different $l$, such that neither $l$ nor $\nexc$ is a good quantum number. However, for the magnetic fields of interest in this work, $l$ and $\nexc$ are good quantum numbers. This is seen in the following way: the diamagnetic term in spherical harmonics \cite{friedrich2017} (Condon-Shortley phase convention)  is
\begin{equation}\label{eq:DiamagneticSymmetry}
    \frac{e^2B^2_0}{8m} (r^2_x+r^2_y)
    =
    \frac{e^2B^2_0\sqrt{\pii}}{6m} r^2 \bigg(Y_{0,0}-\frac{1}{\sqrt{5}}Y_{2,0}\bigg)
\,.
\end{equation}
Since the system is in the Paschen-Back regime, our choice of state $\ket*{\ed}$ entails that this operator only drives transitions with $\Delta{l}=2$. As an example, the energy splitting between the manifold $\ket*{50^2\mathrm{P}_\frac{1}{2}}$ and the energetically closest F states $\ket*{48^2\mathrm{F}_\frac{5}{2}}$ is roughly 65\,GHz (for \srp). The radial quadrupole transition matrix element is roughly $3\times10^6\aB^2$. Therefore, even for $B_0=300\,\mathrm{mT}$, the magnitude of the coupling strength is $e^2B^2_0\sqrt{\pii}\muq\slash(6\sqrt{5}h\me)\approx4.2\,\mathrm{GHz}$, and $l$, as well as $\nexc$, approximately remain good quantum numbers. We finally note that Eq.~\eqref{eq:DiamagneticSymmetry} yields first-order shifts in the Paschen-Back regime, as shown in Fig.~\ref{fig:Transitions}(b). However, these shifts are the same for each ion and only depend on the number of excitons $\nexc$. Hence, we can ignore them, because they only contribute an additive constant to the effective Hamiltonian associated with the subspace $\nexc$. However, we remark that these energy shifts scale as ${\sim}B^2_0$ and are therefore (quadratically) small for weak magnetic fields.

Up to this point, we have expanded the internal Hamiltonian in Eq.~\eqref{eq:ManyBodyInternalHamiltonian} to second order in $\vepse$ and $\vepsv$. Further, we chose in the regime of weak (i) and strong (ii) magnetic fields electronic states $\ket*{\eu}$ and $\ket*{\ed}$, representing the electronic degrees of freedom for each ion. Finally, we verified that with this choice of states, the number of excitons $\nexc$, i.e., the number of ions in the state $\ket*{\eu}$ is (approximately) a good quantum number. However, the perturbations $H_\alpha$, $H_\beta$, and $H_\mathrm{dia}$ can lead to second-order energy shifts. To obtain these energy shifts, we first note that the off-diagonal part of the diamagnetic term only drives far-detuned transitions with $\Delta{l}=2$. Hence, for sufficiently small magnetic fields it does not yield appreciable energy shifts. Therefore, we only consider $H_\alpha$ and $H_\beta$. These can be treated separately, because they have different symmetries and therefore drive different electronic transitions.

We start with $H_\alpha$, which we can approximately treat as a time-independent Hamiltonian, because the \ac*{rf} is far detuned from electronic transitions, yet typically fast compared to the time-scales of interest. Hence, we may average over an \ac*{rf}-cycle, analogously to the treatment in Ref.~\cite{mueller2008}, in order to get an understanding of the qualitative physics this contribution entails. Importantly, the second-order energy shifts associated with $H_\alpha$ are the same for each ion, like the energy shifts introduced by $H_\mathrm{dia}$. Therefore, we can ignore these energy shifts on each energetically isolated subspace with fixed $\nexc$. In contrast, $H_\beta$ results in second-order energy shifts that do depend on the position of the ion in the crystal. The leading-order contribution is [see transition in Fig.~\ref{fig:Transitions}(a)]
\begin{multline}\label{eq:QuadShifts}
    \hbar\omega_z\Delta^{(u)}
    \approx
    (e\beta)^2
    \frac{
        \abs*{\mel*{n^2\mathrm{P}_\frac{3}{2}(\frac{1}{2})}{(r^2-3z^2)}{\ed}}^2
    }{
        E_\ed-E_{\eta}
    }
    \big[q^{(u)}\big]^2
\,,
\end{multline}
with $\eta=(n,1,\frac{1}{2},\frac{3}{2},\frac{1}{2})$. The magnitude of these energy shifts is determined by $(e\beta\muq)^2\slash(E_\ed-E_{\eta})$. For $n=50$, $B_0=20\,\mathrm{mT}$, and \srp ($\beta=10^7\,\mathrm{Vm}^{-2}$) they are on the order of 0.12\,MHz. However, due to their scaling $\beta^2n^{11}$, these energy shifts may become comparable to typical magnitudes of the dipole-dipole coupling strengths in Eq.~\eqref{eq:TransportRates}. In the following we will treat the energy shifts \eqref{eq:QuadShifts} as variables, because they could be additionally engineered via inhomogeneous magnetic fields with amplitudes in the range of $100\,\text{\textmu{T}}\ll{B_0}$, or Stark shifts, both of which can yield detunings of a few to a few tens of megahertz.

In conclusion of this Subsection, we obtain the following simple form for the internal Hamiltonian in Eq.~\eqref{eq:InternalHamiltonian}:
\begin{multline}\label{eq:FinalInternalHamiltonian}
    \frac{H_\mathrm{int}}{\hbar\omega_z}
    =
    \sum^N_{u=1}
    \big[ \Delta^{(u)}_\eu P^{(u)}_\eu
          +
          \Delta^{(u)}_\ed P^{(u)}_\ed \big]
\\
    +
    \sum^N_{u=1}
    \sum^N_{v(\neq{u})=1}
    J^{(u,v)}
    \big[ S^{(u)}_+ S^{(v)}_-
          +
          S^{(u)}_- S^{(v)}_+ \big]
\,,
\end{multline}
with the local projectors $P^{(u)}_\eta=\dyad*{\eta}^{(u)}$ ($\eta=\eu,\ed$) and the tunable detunings $\Delta^{(u)}_\eta$ ($\eta=\eu,\ed$). We remark that Eq.~\eqref{eq:FinalInternalHamiltonian} is valid in both the weak-field (i) and the Paschen-Back regime (ii).

%%% SUBSECTION
\subsection{External-internal-coupling Hamiltonian}

The final goal is to also expand the vibronic contribution in Eq.~\eqref{eq:FullHamiltonian} in the small parameters $\vepse$ and $\vepsv$. It is composed of the single-body terms in Eq.~\eqref{eq:CouplingHamiltonian} and the charge-dipole interaction Hamiltonian in Eq.~\eqref{eq:ChargeDipoleInteraction}:
\begin{equation}\label{eq:ManyBodyCouplingNotSimplified}
    H_\mathrm{int\text{-}ext}(t)
    =
    \sum^N_{u=1} H^{(u)}_\mathrm{int\text{-}ext}(t)
    +
    H_\mathrm{cd}
\,.
\end{equation}
Before we expand this contribution in $\vepsv$ and $\vepse$, we simplify it by using the equilibrium condition \eqref{eq:EquilibriumCondition}, which explicitly reads ($i=x,y,z$)
\begin{equation}
    a^2_i \bar{R}^{(u)}_{0,i}
    =
    \sum^N_{v(\neq{u})=1}
    \frac{\bar{R}^{(u,v)}_{0,i}}
         {\norm*{\bar{\bm{R}}^{(u,v)}_0}^3}
\,.
\end{equation}
By plugging this condition into the charge-dipole interaction Hamiltonian in Eq.~\eqref{eq:ChargeDipoleInteraction} and subsequently expanding to second order, we obtain
\begin{multline}
    H_\mathrm{cd}
    \approx
    4 e \beta
    \sum^N_{u=1}
    \sum_{i=x,y,z}
    a^2_i R^{(u)}_{0,i} r^{(u)}_i
\\
    +
    4 e \beta
    \sum^N_{u=1}
    \sum^N_{v(\neq{u})=1}
    \delta\bm{R}^{(u,v)}
    \cdot
    \frac{S(\bar{\bm{R}}^{(u,v)}_0)}{\norm*{\bar{\bm{R}}^{(u,v)}_0}^3}
    \bm{r}^{(u)}
\,,
\end{multline}
where we defined $\delta\bm{R}^{(u,v)}=\delta\bm{R}^{(u)}-\delta\bm{R}^{(v)}$. Substituting this back into Eq.~\eqref{eq:ManyBodyCouplingNotSimplified} yields
\begin{multline}\label{eq:ExpandedCouplingHamiltonian}
    H_\mathrm{ext\text{-}int}(t)
    =
    2 e \alpha
    \sum^N_{u=1}\sum_{i=x,y}
    r^{(u)}_i
    \big[f_i(t)R^{(u)}_{0,i}+g_i(t)\delta{R}^{(u)}_i\big]
\\
    +
    4 e \beta
    \sum^N_{u=1}
    \bm{r}^{(u)}
    \cdot
    D^{(u)}
    \delta\bm{R}^{(u,v)}
    -
    4 e \beta \sum^N_{u=1} r^{(u)}_z\delta{R}^{(u)}_z
\,,
\end{multline}
where we used Eq.~\eqref{eq:TrapAnisotropies} to simplify terms, introduced the dimensionless functions
\begin{equation}
    f_{x(y)}(t)
    =
    \mp
    \cos(\omgRF t)
    +
    \frac{e\alpha}{M\omgRF^2}
    -
    \frac{M\omgc^2}{8e\alpha}
\,,
\end{equation}
made use of the functions $g_{x(y)}$ [see Eq.~\eqref{eq:gFunctions}], and defined the matrix
\begin{equation}\label{eq:DMatrix}
    D^{(u)}
    =
    \sum^N_{v(\neq{u})=1}
    \frac{S(\bar{\bm{R}}^{(u,v)}_0)}
    {\norm*{\bar{\bm{R}}^{(u,v)}_0}^3}
\,.
\end{equation}
We remark that in Eq.~\eqref{eq:ExpandedCouplingHamiltonian} there are terms of the form $r^{(u)}_iR^{(u)}_{0,i}$ ($i=x,y$). This means that the $\sigma_\pm$ dipole transition matrix elements increase linearly with the radial displacement of the ions at equilibrium, $R^{(u)}_{0,i}$.

We continue by expanding Eq.~\eqref{eq:ExpandedCouplingHamiltonian} in $\vepsv$ and $\vepse$. To this end, we replace the structure matrix with $S(\bm{\bar{R}}^{(u,v)}_0)\approx\mathrm{diag}(1,1,-2)$ and approximate $\norm*{\bm{\bar{R}}^{(u,v)}_0}\approx\abs*{\bar{R}^{(u,v)}_{0,z}}$ in the denominator of Eq.~\eqref{eq:DMatrix}, i.e., we omit third-order contributions in $\vepse$ and $\vepsv$. To see if the number of excitons, $\nexc$, is still a good quantum number, recall that we consider ionic chains, i.e., ion crystals in which the equilibrium displacement in the directions $x$ and $y$ of each ion is in the range of ${\sim}1\text{--}10\xhoz$. Therefore, we estimate the maximal magnitude of the dipole transition matrix element to be $(e\alpha\times10\xhoz\times1000\aB)\slash{h}{\sim}1\,\mathrm{GHz}$ ($\xhoz=10\,\mathrm{nm}$, $\alpha=10^9\,\mathrm{Vm^{-2}}$), which is still much smaller than the energy splitting between S and P, or P and D states. Hence, $\nexc$ is still (approximately) a good quantum number and we may treat Eq.~\eqref{eq:ExpandedCouplingHamiltonian} as a second-order perturbation.

Importantly, we may treat the vibronic Hamiltonian in Eq.~\eqref{eq:ExpandedCouplingHamiltonian} independently of the internal Hamiltonian \eqref{eq:ManyBodyInternalHamiltonian}, when computing second-order energy shifts. This is because these Hamiltonians have different symmetries and therefore are associated with different electronic transitions. To estimate the displacement-dependent energy shifts we note that terms proportional to $\alpha^2$ yield shifts that are typically two orders of magnitude larger than terms of order $\alpha\beta$, and even four orders of magnitude larger than terms proportional to $\beta^2$. This is because, typically, $\alpha\slash\beta{\sim}10^2$ \cite{schmidtkaler2011}. The energy shifts proportional to $\alpha^2$ can be estimated in the following way: as established earlier when we assessed whether $\nexc$ remains a good quantum number, the dipole transition matrix elements proportional to $\alpha$ can be on the order of 1\,GHz. With the energy splitting between different $l$ being on the order of 50\,GHz \cite{higgins2019}, we find that the energy shifts associated with the terms proportional to $\alpha^2$ can be on the order of  ${\sim}(1^2\slash{50})\,\mathrm{GHz}=20\,\mathrm{MHz}$. This shows that they can be comparable to the coupling strengths $J^{(u,v)}$ in Eq.~\eqref{eq:TransportRates}. Further, we observe that energy shifts proportional to $\alpha\beta$ become relevant only for atypically large $\beta$. However, since we are interested in typical parameter regimes, the appreciable energy shifts are those of order $\alpha^2$. Note that this is expected, because Hamiltonian \eqref{eq:ExpandedCouplingHamiltonian} is of order $\vepsv\vepse$ already, such that a second-order perturbation effectively results in fourth-order contributions. However, for the energy shifts proportional to $\alpha^2$, the `smallness' of the fourth-order is compensated by the `largeness' of $\alpha^2$, compared to $\beta^2$.

In conclusion, for Eq.~\eqref{eq:ManyBodyCouplingNotSimplified}, we arrive at the simple expression
\begin{equation}\label{eq:CouplingSecondOrderShifts}
    H_\mathrm{ext\text{-}int}(t)
    \approx
    (2e\alpha)^2
    \sum^N_{u=1}
    \sum_{\eta=\eu,\ed}
    P^{(u)}_\eta
    \sum_{\eta'\neq\eta}
    \frac{\big\vert M^{(u)}_{\eta,\eta'}(t) \big\vert^2}{E_\eta-E_{\eta'}}
\,, \
\end{equation}
with the matrix elements
\begin{multline}
    M^{(u)}_{\eta,\eta'}(t)
    =
    F_x(R^{(u)}_{0,x},\delta{R}^{(u)}_x,t)
    \mel*{\eta}{r_x}{\eta'}
\\
    +
    F_y(R^{(u)}_{0,y},\delta{R}^{(u)}_y,t)
    \mel*{\eta}{r_y}{\eta'}
\,,
\end{multline}
where ($i=x,y$)
\begin{equation}
    F_i(R,\delta{R},t)
    =
    f_i(t)R+g_i(t)\delta{R}
\,.
\end{equation}

Next, we focus on the time-dependence. Again, the \ac{rf} $\nuRF$ is far detuned from any dipole transition in the spectrum, such that no additional transitions are driven by the time-dependence. Therefore, we treat the time-dependence (as before) simply by averaging over an \ac{rf}-cycle: explicitly, we define
\begin{equation}
    \ev*{\phi}_\mathrm{T}
    \equiv
    \nuRF \int^{\frac{1}{\nuRF}}_0 \dd{t} \phi(t)
\,,
\end{equation}
for any function $\phi$ of time. With this definition we find
\begin{multline}\label{eq:CouplingAveragedMatrixElementSquared}
    \ev{\big\vert M^{(u)}_{\eta,\eta'} \big\vert^2}_\mathrm{T}
    =
    \sum_{i=x,y}
    \big\langle
        F^2_i(R^{(u)}_{0,i},\delta{R}^{(u)}_i)
    \big\rangle_\mathrm{T}
    \abs*{\mel*{\eta}{r_i}{\eta'}}^2
\\
    +
    \ev{
        F_x(R^{(u)}_{0,x},\delta{R}^{(u)}_x)
        F_y(R^{(u)}_{0,y},\delta{R}^{(u)}_y)
    }_\mathrm{T}\times
\\
    2\times\mathrm{Re}\big[\mel*{\eta}{r_y}{\eta'}\mel*{\eta}{r_x}{\eta'}^*\big]
\,.
\end{multline}
Interestingly, there is no mixing between the $x$- and the $y$-direction, because from the angular part of the matrix elements,
\begin{multline}
    \big[
        \mel*{\eta}{r_x}{\eta'}
        \mel*{\eta}{r_y}{\eta'}^*
    \big]_\mathrm{ang.}
    =
\\
    (-\ii) \frac{2\pii}{3} r^2
    \mel*{\eta}{\big(Y_{1,-1}-Y_{1,1}\big)}{\eta'}_\mathrm{ang.} \times
\\
    \mel*{\eta}{\big(Y_{1,-1}+Y_{1,1}\big)}{\eta'}^*_\mathrm{ang.}
\,,
\end{multline}
it follows that the final term in Eq.~\eqref{eq:CouplingAveragedMatrixElementSquared} vanishes. This is because the angular matrix elements, which still appear in the expression above, are real. Furthermore, due to isotropy of space, the $x$- and $y$-matrix elements differ only by a phase. Combining all of this reduces Eq.~\eqref{eq:CouplingSecondOrderShifts} to
\begin{equation}\label{eq:TimeAveragedCouplingHamiltonian}
    \ev*{H_\mathrm{ext\text{-}int}}_\mathrm{T}
    =
    \sum^N_{u=1}
    \sum_{i=x,y}
    \big\langle
        F^2_i(R^{(u)}_{0,i},\delta{R}^{(u)}_i)
    \big\rangle_\mathrm{T}
    E^{(u)}
\,,
\end{equation}
with the electronic operators
\begin{equation}
    E^{(u)}
    =
    M \omega^2_z
    \sum_{\eta=\eu,\ed}
    \varepsilon_\eta P^{(u)}_\eta
\end{equation}
and the dimensionless coefficients
\begin{equation}\label{eq:VarepsilonEta}
    \varepsilon_\eta
    =
    \sum_{\eta'\neq\eta}
    \frac{\alpha}{\beta}
    \frac{e\alpha\abs*{\mel*{\eta}{r_x}{\eta'}}^2}{E_\eta-E_{\eta'}}
\,,
\end{equation}
proportional to the dipole polarisability of the Rydberg state $\ket*{\eta}$ ($\eta=\eu,\ed$). Next we calculate the time-average. For a general $R$ and $\delta{R}$ we have ($i=x,y$)
\begin{equation}
    \ev*{F^2_i(R,\delta{R})}_\mathrm{T}
    =
    \ev*{f^2_i}_\mathrm{T}R^2
    +
    \ev*{g^2_i}_\mathrm{T} \delta{R}^2
    +
    2\ev*{f_ig_i}_\mathrm{T} R \delta{R}
\,.
\end{equation}
Explicitly we find
\begin{multline}
    \ev*{F^2_{x(y)}(R,\delta{R})}_\mathrm{T}
    =
    \bigg(
        \bigg[\frac{e\alpha}{M\omgRF^2}-\frac{M\omgc^2}{8e\alpha}\bigg]^2
        +
        \frac{1}{2}
    \bigg) R^2
\\
    +
    \bigg(
        \bigg[\frac{\beta(1\pm\epsilon)}{\alpha}-\frac{M\omgc^2}{4e\alpha}\bigg]^2
        +
        \frac{1}{2}
    \bigg) \delta{R}^2
\\
    +
    2\bigg(\bigg[\frac{e\alpha}{M\omgRF^2}\bigg]^2+\frac{1}{2}\bigg)R\delta{R}
\,.
\end{multline}
For typical values, such as $\alpha=10^9\,\mathrm{Vm^{-2}}$, $\beta=10^7\,\mathrm{Vm^{-2}}$, and $\nuRF=20\,\mathrm{MHz}$ \cite{schmidtkaler2011}, we find $(e\alpha)^2\slash(M\omgRF^2)^2\approx5\times10^{-3}$, and $(\alpha\slash\beta)^2=10^{-4}$. Moreover, even for $B_0=300\,\mathrm{mT}$, $M\omgc^2\slash(e\alpha)\approx10^{-4}$ (for \srp and $\alpha=10^9\,\mathrm{Vm^{-2}}$). Hence, in good approximation,
\begin{equation}
    \ev*{\abs*{F_i(R,\delta{R})}^2}_\mathrm{T}
    \approx
    \frac{(R+\delta{R})^2}{2}
\,.
\end{equation}
Plugging this result back into Eq.~\eqref{eq:TimeAveragedCouplingHamiltonian} yields the time-independent Hamiltonian
\begin{equation}\label{eq:FinalCouplingHamiltonian}
    H_\mathrm{ext\text{-}int}
    \approx
    \frac{1}{2}
    \sum^N_{u=1}
    \big[ \big(R^{(u)}_x\big)^2 + \big(R^{(u)}_y\big)^2 \big]
    E^{(u)}
\,,
\end{equation}
responsible for vibronic coupling.

%%% SUBSECTION
\subsection{Combining vibrational, electronic, and vibronic contributions}

The full Hamiltonian in Eq.~\eqref{eq:FullHamiltonian} is now expanded to second order in $\vepsv$ and $\vepse$, and we found that $\nexc$ is approximately a good quantum number, i.e., the Hamiltonian is approximately U(1)-symmetric. Combining Eq.~\eqref{eq:FinalExternalHamiltonian}, Eq.~\eqref{eq:FinalInternalHamiltonian}, and Eq.~\eqref{eq:FinalCouplingHamiltonian}, according to Eq.~\eqref{eq:FullHamiltonian}, yields
\begin{align}
\begin{split}
    H
    &=
    \frac{\bm{P}^2}{2M}
    +
    \ecz \bar{\bm{Q}}^\mathrm{T} \Gamma^2 \bar{\bm{Q}}
\\
    &+
    \hbar\omega_z
    \sum^N_{u=1}
    \big[ \Delta^{(u)}_\eu P^{(u)}_\eu
          +
          \Delta^{(u)}_\ed P^{(u)}_\ed \big]
\\
    &+
    \hbar\omega
    \sum^N_{u=1}
    \sum^N_{v(\neq{u})=1}
    J^{(u,v)}
    \big[ S^{(u)}_+ S^{(v)}_-
          +
          S^{(u)}_- S^{(v)}_+ \big]
\\
    &+
    \frac{1}{2}
    \sum^N_{u=1}
    \big[ \big(R^{(u)}_x\big)^2 + \big(R^{(u)}_y\big)^2 \big]
    E^{(u)}
\,.
\end{split}
\end{align}
To bring this Hamiltonian to the form of Eq.~\eqref{eq:Model} in the main text, we first define the dimensionless collective modes
\begin{equation}
    \bm{q} = \frac{1}{\xhoz} \bm{Q}
\end{equation}
as well as their canonically conjugate momenta
\begin{equation}
    \bm{p} = \frac{\xhoz}{\hbar} \bm{P}
\,.
\end{equation}
We further define the dimensionless potentials
\begin{equation}\label{eq:PotentialsFromPolarisation}
    W^{(u)}(\bm{q})
    =
    \frac{1}{2}\bm{q}^\mathrm{T}A^{(u)}\bm{q}
    -
    \bm{f}^{(u)}\bm{q}
\,,
\end{equation}
with the curvatures
\begin{equation}\label{eq:DisplacementsStructureMatrices}
    A^{(u)}_{\mu,\nu}
    =
    O^\mathrm{T}_{\mu,(u,x)}O_{(u,x),\nu}+O^\mathrm{T}_{\mu,(u,y)}O_{(u,y),\nu}
\,
\end{equation}
and the mechanical forces
\begin{equation}
    f^{(u)}_\mu
    =
    -\tilde{R}^{(u)}_{0,x} O_{(u,x),\mu}-\tilde{R}^{(u)}_{0,y} O_{(u,y),\mu}
\,,
\end{equation}
where $O$ is the rotation introduced in Subsec.~\ref{subsec:ExternalHamiltonian}. Further, we set $\tilde{\xi}=\xi\slash\xhoz$, for any quantity $\xi$, and introduced the shorthand notation $(u,i)=3(u-1)+i$, for $i=x,y,z$ (identified with $i=1,2,3$). Finally, we defined the geometrically induced detunings
\begin{equation}\label{eq:GeometricDetunings}
    \Delta^{(u)}_{0,\eta}
    =
    \frac{\varepsilon_\eta}{2}
    \big[ \big(\tilde{R}^{(u)}_{0,x}\big)^2
          +
          \big(\tilde{R}^{(u)}_{0,y}\big)^2 \big]
\,.
\end{equation}
Expressed in the newly defined quantities, the full Hamiltonian takes the form
\begin{align}\label{eq:ModelSIUnits}
\begin{split}
    \frac{H}{\hbar\omega_z}
    &=
    \frac{1}{2}\big[\bm{p}^2 + \bm{q}^\mathrm{T}\Gamma^2\bm{q}\big]
\\
    &+
    \sum^N_{u=1}
    \big[ \delta^{(u)}_\eu P^{(u)}_\eu
          +
          \delta^{(u)}_\ed P^{(u)}_\ed
    \big]
\\
    &+
    \sum^N_{u=1}
    \sum^N_{v(\neq{u})=1}
    J^{(u,v)}
    \big[ S^{(u)}_+ S^{(v)}_-
          +
          S^{(u)}_- S^{(v)}_+ \big]
\\
    &+
    \sum^N_{u=1}
    \big[ W^{(u)}(\bm{q}) \varepsilon_\eu P^{(u)}_\eu
          +
          W^{(u)}(\bm{q}) \varepsilon_\ed P^{(u)}_\ed \big]
\,,
\end{split}
\end{align}
where we introduced the shorthand $\delta^{(u)}_\eta=\Delta^{(u)}_{0,\eta}+\Delta^{(u)}_\eta$ for the detunings. From this result, Eq.~\eqref{eq:Model} from the main text is recovered upon the substitution $\hbar\omega_z\mapsto1$.

%%% FIGURE
\figElectronicParameters

For completeness, Fig.~\ref{fig:ElectronicParameters} shows values for the dimensionless parameter $\varepsilon_\eta$ as a function of the principal quantum number $n$, as well as the detunings $\Delta^{(u)}_{0,\eta}$ as a function of the displacement of an ion from the trap axis (here, the $z$-axis).

Finally, we remark that for magnetic fields in the range where one is neither in the weak-field Zeeman, nor in the Paschen-Back regime, Fig.~\ref{fig:ElectronicParameters} assumes an interpolated form between panels (a) [(b)] and (c) [(d)]. For simplicity and concreteness, we assume that the system is in the Paschen-Back regime for all further discussions in this work: all results are still obtainable in the weak-field Zeeman regime.

%%% SUBSECTION
\subsection{Interpretation of state-dependent potentials}

In this final Subsection we remark on an important aspect concerning the Hamiltonian in Eq.~\eqref{eq:ModelSIUnits}: generally, Eq.~\eqref{eq:ModelSIUnits} only captures the short-time dynamics after electronic excitation. To understand this, note the following: in general, the electronic excitation of an ion crystal shifts its equilibrium position. However, according to the Franck-Condon principle \cite{franck1926,condon1928, atkins2011}, which states that the conformation of a molecule is unchanged during electronic excitation, the geometry of the ion crystal is not at equilibrium immediately after the excitation. Therefore, the crystal experiences a force, which displaces it towards the shifted equilibrium. To illustrate this, we consider a zig-zag ion crystal with $N=3$, which undergoes a transition from the ground state $\ket*{\mathrm{GGG}}$ to the P manifold $\ket*{\ed\ed\ed}$. In Fig.~\ref{fig:RelevanceOfQuarticTerm}(a),(b) we show a section through the full molecular potential \eqref{eq:StateDependentPotential} along the curve $\gamma(\tilde{x})=\bm{R}^{\mathrm{G})}_0+\xhoz(\tilde{x}-\tilde{x}_0)[-1,0,2,0,-1,0]^\mathrm{T}$. Here, $\bm{R}^{\mathrm{G})}_0$ is the equilibrium position associated with the electronic ground state $\ket*{\mathrm{GGG}}$ and $\tilde{x}^{(\mathrm{G})}_0$ is the value of the reaction coordinate $\tilde{x}$ at equilibrium [see inset in panel (a)]. The curve shows two minima, which correspond to the zig-zag and zag-zig conformation. Note that the grey line indicates the equilibrium position of the molecular potential in the electronic ground state $\ket*{\mathrm{GGG}}$. Importantly, after the electronic excitation from $\ket*{\mathrm{GGG}}$ to the P manifold $\ket*{\ed\ed\ed}$ the ion crystal is still located at this position, which is not the minimum, as the black curve in Fig.~\ref{fig:RelevanceOfQuarticTerm}(a),(b) shows. Therefore, the crystal experiences a force. Importantly, the parabola (red dashed line) approximates the potential close to the ground-state equilibrium position (grey line). However, since the shift of the equilibrium upon electronic excitation is small, it still approximates the minimum of the molecular potential. Because of this, the resetting force introduced by the left arm of the parabola approximates the true resetting force. However, when the shift of the equilibrium position upon electronic excitation is large, this is not the case. This scenario is shown in Fig.~\ref{fig:RelevanceOfQuarticTerm}(c),(d), which shows an analogous molecular potential but with a larger polarisability of the Rydberg P state. Here, the minimum of the black curve is not at all captured by the harmonic approximation (red dashed parabola). In this case, the harmonic approximation used in Eq.~\eqref{eq:ModelSIUnits} (red dashed line) predicts a resetting force even when the molecular potential (black curve) does not yield one. This is because the molecular potential is not harmonically approximated at a minimum. As a consequence, for too long evolution after the electronic excitation, the ionic crystal is displaced far from the region where the harmonic approximation is valid.

%%% FIGURE
\figRelevanceOfQuarticTerm

Generally, to obtain the relevant harmonic approximation of the molecular potential close to some geometry $\bm{R}_0$, one must consider the full electronic state-dependent molecular potential
\begin{multline}\label{eq:StateDependentPotential}
    \frac{V_\mathrm{ext}(\bm{R})}{\ecz}
    =
    \sum^N_{u=1}\sum^{N}_{v(\neq{u})=1}
    \frac{1}{\norm*{\bar{\bm{R}}^{(u,v)}}}
\\
    +
    \sum^N_{u=1}\sum_{\eta=\eu,\ed,\mathrm{G}}
    \bigg[
    \sum_{i=x,y}
    (a^2_i+\varepsilon_\eta)
    \big(\bar{R}^{(u)}_i\big)^2
    +
    \big(\bar{R}^{(u)}_z\big)^2
    \bigg]
    P^{(u)}_\eta
\,,
\end{multline}
which results from combining Eq.~\eqref{eq:GroundStateExternalfPotential} with Eq.~\eqref{eq:FinalCouplingHamiltonian}. Note that we included the electronic ground state $\ket*{\mathrm{G}}$ in the sum, for which $\varepsilon_\mathrm{G}\approx0$, such that Eq.~\eqref{eq:GroundStateExternalfPotential} is recovered for an ion crystal in the electronic ground state. From this potential, one then calculates the state-dependent equilibrium geometry with the condition that the state-dependent potential gradient (in units of $\ecz\slash\ell_z$) must vanish:
\begin{multline}
    \frac{\nabla^\mathrm{T}_{\bar{\bm{R}}}V_\mathrm{ext}(\bm{R}_0)}{\ecz}
    =
    2\bigoplus^N_{u=1}
    \bigg(
        \sum_{\eta=\eu,\ed,\mathrm{G}}
        \mqty[(a^2_x+\varepsilon_\eta) \bar{R}^{(u)}_{0,x} \\
              (a^2_y+\varepsilon_\eta) \bar{R}^{(u)}_{0,y} \\
              \bar{R}^{(u)}_{0,z}]
        P^{(u)}_\eta
\\
    -
    \sum^N_{v(\neq{u})=1}
    \frac{\bar{\bm{R}}^{(u,v)}_0}
         {\norm*{\bar{\bm{R}}^{(u,v)}_0}^3}
    \bigg)
    =
    0
\,.
\end{multline}
At the equilibrium position $\bm{R}_0$ the Hessian matrix (in units of $M\omega^2_z$) [cf. Eq.~\eqref{eq:ApproximatedPotential}] takes the form
\begin{multline}\label{eq:GeneralHessian}
    K(\bm{R}_0)
    =
    \bigoplus^N_{u=1}\sum_{\eta=\eu,\ed,\mathrm{G}}
    \mqty[a_x^2+\varepsilon_\eta & & \\ & a^2_y+\varepsilon_\eta & \\ & & 1]
    P^{(u)}_\eta
\\
    +
    \frac{1}{2}
    \sum^N_{u=1}\sum^N_{v(\neq{u})=1}
    \frac{C^{(u,v)} \otimes S(\bar{\bm{R}}^{(u,v)}_0)}
         {\norm*{\bar{\bm{R}}^{(u,v)}_0}^3}
\,,
\end{multline}
with the structure matrices defined in Eq.~\eqref{eq:StructureMatrix} and the matrices $C^{(u,v)}_{i,j}=\delta_{u,i}\delta_{v,j}+\delta_{v,i}\delta_{u,j}-\delta_{u,i}\delta_{u,j}-\delta_{v,i}\delta_{v,j}$. With the general analytical expressions for the potential gradient and the Hessian given above, it is straightforward to compute the collective-mode frequencies and mechanical forces for any electronic state at any desired geometry $\bm{R}_0$.

%%% SECTION
\section{Hamiltonian for an ion crystal consisting of three ions}
\label{app:DerivationOfTheThreeIonHamiltonian}

Here, we derive Eq.~\eqref{eq:ThreeIonModel} from the main text. It describes an ion crystal composed of three ions immediately after Franck-Condon excitation \cite{franck1926,condon1928,atkins2011} from the equilibrium geometry of an ion crystal in its electronic ground state $\ket*{\bm{\mathrm{G}}}\equiv\ket*{\mathrm{GGG}}$ to the molecular potentials associated with the subspace $\nexc=1$. To simplify the problem for the analytical discussion, we `freeze-out' the $y$-direction in this work and only consider collective modes displacing the ionic crystal in $x$- and $z$-direction. Further, to keep the discussion simple, we still employ Eq.~\eqref{eq:TrapAnisotropies} with $\epsilon=0$ to determine the trap anisotropy in $x$-direction. With these approximations, the vibrational dynamics is described by six collective modes, where two of them describe the centre-of-mass motion. To analytically calculate the remaining four collective modes and their associated frequencies, we first determine the equilibrium geometry of the ion crystal in the electronic ground state $\ket*{\bm{\mathrm{G}}}$. To this end, we use the symmetry relations $R^{(1)}_{0,x}=R^{(3)}_{0,x}$, $R^{(2)}_{0,x}=-2R^{(1)}_{0,x}$, $R^{(1)}_{0,z}=-R^{(3)}_{0,z}$, and $R^{(2)}_{0,z}=0$, which hold true for the equilibrium position $\bm{R}_0$, i.e., $\bm{R}_0$ obeys Eq.~\eqref{eq:EquilibriumCondition}. Using these relations, we analytically calculate the equilibrium values $x_0=-R^{(1)}_{0,x}$, and $z_0=R^{(3)}_{0,z}$, for a fixed trap-anisotropy $a_x$. We distinguish two cases: for $a^2_x>12\slash{5}$ the system is in the linear phase ($x_0=0$), with $z_0=(5\slash{4})^{1\slash{3}}$. In contrast, for $a^2_x<12\slash{5}$ the system is in the zig-zag phase. In this case we obtain
\begin{align}
    x_0
    =
    \frac{1}{3}
    \sqrt{\bigg(\frac{3}{a^2_x}\bigg)^{\frac{2}{3}}
          -
          z^2_0
         }
\,, \
    z_0
    =
    \bigg[4\bigg(1-\frac{a^2_x}{3}\bigg)\bigg]^{-\frac{1}{3}}
\,.
\end{align}
Note that $-x_0$ is also a solution; it corresponds to the zig-zag 
geometry reflected at the $z$-axis (`zag-zig'). Note that $x_0=x^{(\mathrm{G})}_0$ in the notation used in Fig.~\ref{fig:RelevanceOfQuarticTerm}. Using these quantities, the Hessian in both the linear and the zig-zag phases can be expressed according to the general formula \eqref{eq:GeneralHessian}, which results in
\begin{widetext}
\begin{equation}\label{eq:GeneralThreeIonHessian}
    K
    =
    \mqty[a^2_x+a-d & c & -a & -c & d & 0 \\
          c & 1+b+2d & -c & -b & 0 & -2d\\
          -a & -c & a^2_x+2a & 0 & -a & c \\
          -c & -b & 0 & 1+2b & c & -b \\
          d & 0 & -a & c & a^2_x+a-d & -c \\
          0 & -2d & c & -b & -c & 1+b+2d \\]
\,,
\end{equation}
\end{widetext}
with the quantities
\begin{equation}
    a = \frac{-1+2\mu^2}{\lambda z^3_0}
\,, \
    b = \frac{2-\mu^2}{\lambda z^3_0}
\,, \
    c = \frac{3\mu}{\lambda z^3_0}
\,, \
    d = \frac{1}{8 z^3_0}
\,,
\end{equation}
where $\mu=3x_0\slash{z_0}$ and $\lambda=(1+\mu^2)^{5\slash{2}}$. Note that in the linear phase $\mu=0$, $\lambda=1$, and $c=0$, such that the Hessian splits into two decoupled three-dimensional blocks (upon reordering). These blocks correspond to the three modes displacing the ion crystal in $x$-direction and the three modes displacing it in $z$-direction, respectively. In this case, the rotation (see Subsec.~\ref{subsec:ExternalHamiltonian})
\begin{equation}\label{eq:SixDimensionalEigenmodesLinearChain}
    O_\mathrm{lin}
    =
    \mqty[ \frac{1}{\sqrt{6}} & \frac{1}{\sqrt{3}} & \frac{1}{\sqrt{2}} & 0 & 0 & 0 \\
           0 & 0 & 0 & \frac{1}{\sqrt{6}} & \frac{1}{\sqrt{3}} & \frac{1}{\sqrt{2}} \\
          -\frac{2}{\sqrt{6}} & \frac{1}{\sqrt{3}} & 0 & 0 & 0 & 0\\
           0 & 0 & 0 & -\frac{2}{\sqrt{6}} & \frac{1}{\sqrt{3}} & 0 \\
          \frac{1}{\sqrt{6}} & \frac{1}{\sqrt{3}} & -\frac{1}{\sqrt{2}} & 0 & 0 & 0 \\
          0 & 0 & 0 & \frac{1}{\sqrt{6}} & \frac{1}{\sqrt{3}} & -\frac{1}{\sqrt{2}} ]
\end{equation}
diagonalises the Hessian. The columns of this matrix define the collective eigenmodes $\bm{q}_\mathrm{lin}=O^\mathrm{T}_\mathrm{lin}\delta\bm{R}$ (see Subsec.~\ref{subsec:ExternalHamiltonian}), where we defined $\bm{q}_\mathrm{lin}^\mathrm{T}=[q^x_1,q^x_2,q^x_3,q^z_1,q^z_2,q^z_3]$. The first three entries are the collective modes shown in Fig.~\ref{fig:MainPicture}(d) in the main text, and the last three are the analogous ones in $z$-direction. For the associated squared collective-mode frequencies we find (in units of $\omega^2_z$)
\begin{align}
\begin{split}
    \Gamma^2_{1,x}
    &=
    a^2_x-\frac{12}{5}
\,, \
    \Gamma^2_{2,x}
    =
    a^2_x
\,, \
    \Gamma^2_{3,x}
    =
    a^2_x-1
\,, \\
    \Gamma^2_{1,z}
    &=
    \frac{29}{5}
\,, \
    \Gamma^2_{2,z}
    =
    1
\,, \
    \Gamma^2_{3,z}
    =
    3
\,.
\end{split}
\end{align}
Note that all of these frequencies are either constant or linear in $a^2_x$.

Next, we consider the ion crystal in the zig-zag phase, i.e., $a^2_x<12\slash{5}$. The first step is to pre-diagonalise the Hessian \eqref{eq:GeneralThreeIonHessian} with the rotation \eqref{eq:SixDimensionalEigenmodesLinearChain}, yielding
\begin{equation}\label{eq:PrediagonalisedHessian}
    K'
    =
    \mqty[a^2_x+3a & 0 & 0 & 0 & 0 & \sqrt{3}c \\
          0 & a^2_x & 0 & 0 & 0 & 0 \\
          0 & 0 & a^2_x+a-2d & \sqrt{3}c & 0 & 0 \\
          0 & 0 & \sqrt{3}c & 3b+1 & 0 & 0 \\
          0 & 0 & 0 & 0 & 1 & 0 \\
          \sqrt{3}c & 0 & 0 & 0 & 0 & b+4d+1
          ]
\end{equation}
for the transformed Hessian $K'=O^\mathrm{T}_\mathrm{lin}KO_\mathrm{lin}$. Reordering the rows and columns of $O_\mathrm{lin}$ shows that $K'$ consists of two one-dimensional and two two-dimensional decoupled blocks. More precisely, we observe that the mode $q_{1,x}$ is coupled to $q_{3,z}$ and $q_{3,x}$ is coupled to $q_{1,z}$. According to Eq.~\eqref{eq:PrediagonalisedHessian} the two two-dimensional blocks are
\begin{align}
    K_1
    &=
    \mqty[a^2_x+3a & \sqrt{3}c \\
          \sqrt{3}c & b+4d+1]
\,, \\
    K_2
    &=
    \mqty[a^2_x+a-2d & \sqrt{3}c \\
          \sqrt{3}c & 3b+1]
\,.
\end{align}
Note that in the linear phase $\mu=0$ such that $c=0$ and $K_1$ and $K_2$ become diagonal. We diagonalise $K_1$ and $K_2$ with the rotations $O_1$ and $O_2$, respectively, which read
\begin{equation}
    O_\xi
    =
    \mqty[\frac{\sqrt{3}c}{\sqrt{3c^2+\lambda^2_{\xi,-}}}
          & \frac{-\sqrt{3}c}{\sqrt{3c^2+\lambda^2_{\xi,+}}} \\
          \frac{-\lambda_{\xi,-}}{\sqrt{3c^2+\lambda^2_{\xi,-}}}
          & \frac{\lambda_{\xi,+}}{\sqrt{3c^2+\lambda^2_{\xi,+}}} ]
\,,
\end{equation}
where $\xi=1,2$ and
\begin{align}
    \lambda_{1,\pm}
    &=
    \Gamma^2_{1,\pm}+\frac{\mu^2a^2_x}{1+\mu^2}-3
\,, \\
    \lambda_{2,\pm}
    &=
    \Gamma^2_{2,\pm}-\frac{2-\mu^2}{1+\mu^2}a^2_x-1
\,.
\end{align}
For the squared collective-mode frequencies (in units of $\omega^2_z$) we find
\begin{align}
\begin{split}
    \Gamma^2_{1,\pm}
    &=
    \frac{3}{2}+\frac{\mu^2a^2_x}{1+\mu^2}
    \mp
    \sqrt{\frac{\mu^2a^2_x}{1+\mu^2}
          \bigg[\frac{(3+4\mu^2)a^2_x}{1+\mu^2}-6\bigg]
          +\frac{9}{4}}
\,, \\
    \Gamma^2_{2,\pm}
    &=
    \frac{3+\mu^2}{1+\mu^2}\frac{a^2_x}{2}
    \mp
    \sqrt{1
          +
          \bigg[\frac{1+3\mu^2}{1+\mu^2}\bigg]^2\frac{a^4_x}{4}
          +
          \frac{[1-3\mu^2]a^2_x}{1+\mu^2}}
\,.
\end{split}
\end{align}
To interpret this result, note that $\Gamma_{1,+(-)}\to\Gamma_{1,x(3,z)}$ and $\Gamma_{2,+(-)}\to\Gamma_{3,x(1,z)}$ as $\mu\to0$, i.e., in the limit of a linear geometry. Of course, the eigenfrequencies of the modes that constitute the centre-of-mass shifts in $x$- and $z$-direction are unchanged, with respect to the linear geometry. From the block form of the Hessian matrix for linear ion crystals, it follows that close to the zig-zag-to-linear phase transition \cite{fishman2008} we can approximate the collective modes to be the same as for the linear chain. This is because all terms in the external Hamiltonian are already of second order in the expansion parameter $\vepsv^2$ [see Sec.~\ref{subsec:ExternalHamiltonian}], such that for $x_0\slash{z_0}{\sim}\vepsv$ we can ignore corrections to the collective modes generated by the zig-zag geometry. Within this approximation we have
\begin{equation}\label{eq:SumOverAs}
    A^{(1)}+A^{(2)}+A^{(3)}
    =
    1 + \mathcal{O}(\vepsv)
\,,
\end{equation}
for the structure matrices in Eq.~\eqref{eq:PotentialsFromPolarisation}. Therefore, in both the linear and the zig-zag phase, only the three collective modes $\bm{q}=[q^x_1,q^x_2,q^x_3]^\mathrm{T}$ are relevant. Explicitly, these are defined as $\bm{q}=O^\mathrm{T}\delta\bm{R}_x$, with $\delta\bm{R}_x=[\delta{R}^{(1)}_x,\delta{R}^{(2)}_x,\delta{R}^{(3)}_x]$ and
\begin{equation}\label{eq:LinearThreeIonCrystalCollectiveModes}
    O
    =
    \mqty[\frac{1}{\sqrt{6}} & \frac{1}{\sqrt{3}} & \frac{1}{\sqrt{2}} \\
          -\frac{2}{\sqrt{6}} & \frac{1}{\sqrt{3}} & 0 \\
          \frac{1}{\sqrt{6}} & \frac{1}{\sqrt{3}} & -\frac{1}{\sqrt{2}}]
\,.
\end{equation}

The next step is to finally obtain Eq.~\eqref{eq:ThreeIonModel} from the main text. To this end, we apply the approximation \eqref{eq:SumOverAs} to the full Hamiltonian in Eq.~\eqref{eq:FullHamiltonian}. This yields
\begin{align}\label{eq:HamiltonianThreeIonGeneral}
\begin{split}
    \frac{H}{\hbar\omega_z}
    &=
    \frac{1}{2}\big[\bm{p}^2 + \bm{q}^\mathrm{T}\Gamma^2\bm{q}\big]
\\
    &+
    \sum^3_{u=1}
    \big[ \delta^{(u)}_\eu P^{(u)}_\eu
          +
          \delta^{(u)}_\ed P^{(u)}_\ed
    \big]
\\
    &+
    \sum^3_{u=1}
    \sum^3_{v(\neq{u})=1}
    J^{(u,v)}
    \big[ S^{(u)}_+ S^{(v)}_-
          +
          S^{(u)}_- S^{(v)}_+ \big]
\\
    &+
    \sum^3_{u=1}
    \big[ W^{(u)}(\bm{q}) \varepsilon_\eu P^{(u)}_\eu
          +
          W^{(u)}(\bm{q}) \varepsilon_\ed P^{(u)}_\ed \big]
\,.
\end{split}
\end{align}
Here, $\bm{q}^\mathrm{T}=[q_1,q_2,q_3]$, where we introduced the shorthand $q_i\equiv{q}^x_i$ ($i=1,2,3$).

The next step is to express the second term in Hamiltonian \eqref{eq:HamiltonianThreeIonGeneral} in the basis states $\ket*{k}$, $k=1,2,3$, which span the single-excitation subspace visualised in Fig.~\ref{fig:MainPicture}(c) in the main text. As a result, we obtain
\begin{multline}
    \sum^3_{u=1}
    \big[ \delta^{(u)}_\eu P^{(u)}_\eu
          +
          \delta^{(u)}_\ed P^{(u)}_\ed
    \big]
\\
    =
    \delta^{(1)}_\ed+\delta^{(2)}_\ed+\delta^{(3)}_\ed
    +
    \sum^3_{k=1}
    \big(\delta^{(k)}_\eu-\delta^{(k)}_\ed\big) \dyad*{k}
\,.
\end{multline}
The first term can be ignored, because it constitutes an additive constant to the Hamiltonian. Further, we subtract the constant energy shift $\delta^{(1)}_\eu-\delta^{(1)}_\ed$. Up to the subtracted constant energy shifts, we obtain
\begin{multline}\label{eq:ThreeIonSimpl1}
    \sum^3_{u=1}
    \big[ \delta^{(u)}_\eu P^{(u)}_\eu
          +
          \delta^{(u)}_\ed P^{(u)}_\ed
    \big]
\\
    =
    \Delta_0 \dyad*{2}
    +
    \Delta^{(2)} \dyad*{2}
    +
    \Delta^{(3)} \dyad*{3}
\,.
\end{multline}
where
\begin{equation}
    \Delta_0
    \equiv
    \frac{3}{4} \big( \Delta^{(2)}_{0,\eu} - \Delta^{(2)}_{0,\ed} \big)
\end{equation}
and ($k=2,3$)
\begin{equation}
    \Delta^{(k)}
    \equiv
    \Delta^{(k)}_\eu - \Delta^{(k)}_\ed
    -
    \Delta^{(1)}_\eu + \Delta^{(1)}_\ed
\,.
\end{equation}

Next, we express the third term in Eq.~\eqref{eq:HamiltonianThreeIonGeneral} in the basis states $\ket*{k}$. This results in
\begin{multline}\label{eq:ThreeIonSimpl2}
    \sum^3_{u=1}
    \sum^3_{v(\neq{u})=1}
    J^{(u,v)}
    \big[ S^{(u)}_+ S^{(v)}_-
          +
          S^{(u)}_- S^{(v)}_+ \big]
\\
    =
    J
    \sum_{k>l}
    \frac{\dyad*{k}{l}+\dyad*{l}{k}}{\abs*{k-l}^3}
\,.
\end{multline}
To obtain this expression we defined $J=J^{(2,1)}$ and used the inverse cubic scaling of the transport rates $J^{(u,v)}$ stated in Eq.~\eqref{eq:TransportRates}.

The last step is to express the final term in Eq.~\eqref{eq:HamiltonianThreeIonGeneral} in the basis states $\ket*{k}$ as well. This simply results in
\begin{multline}\label{eq:ThreeIonSimpl3}
    \sum^3_{u=1}
    \big[ W^{(u)}(\bm{q}) \varepsilon_\eu P^{(u)}_\eu
          +
          W^{(u)}(\bm{q}) \varepsilon_\ed P^{(u)}_\ed \big]
    =
\\
    \sum^3_{k=1}
    \bigg[
        \varepsilon_\eu W^{(k)}(\bm{q})
        +
        \varepsilon_\ed \sum_{l\neq{k}} W^{(l)}(\bm{q})
    \bigg]
    \dyad*{k}
\,.
\end{multline}

By combining Eqs.~\eqref{eq:ThreeIonSimpl1}, \eqref{eq:ThreeIonSimpl2}, \eqref{eq:ThreeIonSimpl3}, the full Hamiltonian in Eq.~\eqref{eq:HamiltonianThreeIonGeneral} takes the form
\begin{align}
\begin{split}
    \frac{H}{\hbar\omega_z}
    &=
    \sum^3_{k=1} \frac{1}{2}\big[\bm{p}^2 + V^{(k)}(\bm{q})\big] \dyad*{k}
    +
    J
    \sum_{k>l}
    \frac{\dyad*{k}{l}+\dyad*{l}{k}}{\abs*{k-l}^3}
\\
    &+
    \big(\Delta_0+\Delta^{(2)}\big) \dyad*{2} +\Delta^{(3)} \dyad*{3}
\,,
\end{split}
\end{align}
where we introduced the potentials
\begin{equation}
    V^{(k)}(\bm{q})
    =
    \bm{q}^\mathrm{T}\Gamma^2\bm{q}
    +
    \varepsilon_\eu W^{(k)}(\bm{q})
    +
    \varepsilon_\ed \sum_{l\neq{k}} W^{(l)}(\bm{q})
\,.
\end{equation}
To recover Eq.~\eqref{eq:ThreeIonModel} from the main text, we set $\hbar\omega_z\mapsto1$ and $\Delta_0=0$. The latter is justified because in the scenarios presented in the main text we consider linear ion crystals.

%%% SECTION
\section{Minimal model and phonon facilitation}
\label{app:MinimalModelAndPhononFacilitation}

To arrive at Eq.~\eqref{eq:MinimalModel} from the main text, we start with Hamiltonian \eqref{eq:ThreeIonModel} from the main text. In the weak-coupling regime, i.e., where $\varepsilon_\eta$ is small, the vibrational state of the system is only weakly affected as time evolves. Therefore, if we choose the motional coherent state $\ket*{\alpha_\mathrm{cs},0,0}$ (see main text) as the initial state, we can model the system classically. To this end, we first obtain the initial condition for the classical equations of motion defined by the initial state. For this we calculate
\begin{equation}
    \mel*{\alpha_\mathrm{cs}}{q_1}{\alpha_\mathrm{cs}}
    =
    \sqrt{\frac{2}{\sqrt{\Gamma^2_1+\varepsilon_\ed}}}\alpha_\mathrm{cs}
\,.
\end{equation}
Therefore, the appropriate initial condition is $q_1(0)=\sqrt{2\slash\sqrt{\Gamma^2_1+\varepsilon_\ed}}\alpha_\mathrm{cs}$. With the initial condition for the momentum, $p_1(0)=0$, we can solve the classical equations of motion. This results in
\begin{align}
    q_1(t)
    &=
    q_1(0)
    \cos(\sqrt{\Gamma^2_1+\varepsilon_\ed}t)
\,, \\
    p_1(t)
    &=
    -\sqrt{\Gamma^2_1+\varepsilon_\ed} q_1(0)
    \sin(\sqrt{\Gamma^2_1+\varepsilon_\ed}t)
\,.
\end{align}

The next step is to substitute this result into Hamiltonian \eqref{eq:ThreeIonModel} from the main text. To this end, we first rewrite the Hamiltonian describing the ionic motion in the molecular potential surfaces $V^{(k)}$ in a slightly different form:
\begin{equation}\label{eq:IonicMotionLinearThreeIonChain}
    \frac{\bm{p}^2}{2}
    +
    V^{(k)}(\bm{q})
    =
    \frac{1}{2}
    \bm{q}^\mathrm{T}
    \big(\Gamma^2+\varepsilon_\ed)
    \bm{q}
    +
    (\varepsilon_\eu-\varepsilon_\ed)
    W^{(k)}(\bm{q})
\,.
\end{equation}
To arrive at this expression we used Eq.~\eqref{eq:SumOverAs}, which is exact in our case of a linear ion crystal. After substituting the classical solution, the first term in Eq.~\eqref{eq:IonicMotionLinearThreeIonChain} becomes
\begin{equation}
    \frac{1}{2}
    \big[
        \bm{p}^2
        +
        \bm{q}^\mathrm{T}\big(\Gamma^2+\varepsilon_\ed\big)\bm{q}
    \big]
    =
    \alpha^2_\mathrm{cs}\big(\Gamma_1+\varepsilon_\ed\big)
\,.
\end{equation}
This represents only a constant energy shift (due to energy conservation) to the Hamiltonian, and as such, can be disregarded. Next, we express the state-dependent part of Eq.~\eqref{eq:IonicMotionLinearThreeIonChain} in the classical solution:
\begin{equation}\label{eq:StateDependentPartClassical}
    W^{(k)}(\bm{q}(t))
    =
    \frac{A^{(k)}_{11}\alpha^2_\mathrm{cs}}{2\sqrt{\Gamma^2_1+\varepsilon_\ed}}
    \bigg[1+\cos(2\sqrt{\Gamma^2_1+\varepsilon_\ed}t)\bigg]
\,.
\end{equation}
Here, $A^{(k)}_{11}$ is the first component of the matrix $A^{(k)}$ for a linear geometry, defined in Eq.~\eqref{eq:DisplacementsStructureMatrices}, i.e., $A^{(1)}=A^{(3)}=1\slash6$, $A^{(2)}=2\slash3$. The constant term in Eq.~\eqref{eq:StateDependentPartClassical} yields a site-dependent energy shift. In contrast, the time-dependent contribution drives two-phonon transitions with energy separation $2\sqrt{\Gamma^2_1+\varepsilon_\ed}$. However, for weak coupling and a highly occupied initial state, the phonon-mode occupation is approximately constant --- in this case we can omit the time-dependent contribution. Within these approximations we find that Eq.~\eqref{eq:IonicMotionLinearThreeIonChain} reduces to (up to a constant energy shift)
\begin{equation}
    \frac{\bm{p}^2}{2}
    +
    V^{(k)}(\bm{q})
    \approx
    \big(A^{(k)}_{11}-A^{(1)}_{11}\big)
    \frac{\alpha^2_\mathrm{cs}}{2}
    \frac{\varepsilon_\eu-\varepsilon_\ed}
         {\sqrt{\Gamma^2_1+\varepsilon_\ed}}
\,.
\end{equation}
Note that, due to the symmetry of the linear ion crystal, we have $A^{(2)}_{11}-A^{(1)}_{11}=1\slash{2}$ and $A^{(3)}_{11}-A^{(1)}_{11}=0$.

Next, we substitute this result back into Eq.~\eqref{eq:ThreeIonModel} from the main text, and further, for simplicity, omit next-to-nearest-neighbour hopping. This results in the effective Hamiltonian
\begin{multline}
    \frac{H_\mathrm{eff}}{\hbar\omega_z}
    =
    J(\dyad*{1}{2}+\dyad*{2}{3}+\mathrm{h.\,c.})
\\
    +
    \bigg(
        \Delta^{(2)}
        +
        \frac{\alpha^2_\mathrm{cs}}{4}
        \frac{\varepsilon_\eu-\varepsilon_\ed}
             {\sqrt{\Gamma^2_1+\varepsilon_\ed}}
    \bigg)
    \dyad*{2}
    +
    \Delta^{(3)}
    \dyad*{3}
\,,
\end{multline}
where $J=J^{(2,1)}$ is given in Eq.~\eqref{eq:TransportRates}. Finally, for $\Delta^{(2)}=\Delta^{(3)}=0$ the effective Hamiltonian \eqref{eq:MinimalModel} from the main text is recovered.

Note that the detuning resulting from the molecular vibrations can be interpreted as a potential barrier for the exciton. Importantly, the fact that only ion 2 experiences such a potential barrier is not a coincidence: it is a result of the displacement signature of the collective mode $q_1$. Generally, the position of the potential barrier and its strength are controlled by the displacement signature of the initially `kicked' vibrational mode.

As a final remark, we note that the potential barrier resulting from the molecular vibrations can also be used to facilitate the exciton transport. This can happen if the detunings $\Delta^{(2)}$ and $\Delta^{(3)}$ are non-zero. In this case, the potential barrier introduced through the molecular vibrations may cancel existing detunings. To demonstrate this, we show in Fig.~\ref{fig:Facilitation}(a) the same simulation of the exciton densities as in Fig.~\ref{fig:Blockage}(b) from the main text. However, here we use $\Delta^{(2)}=-(\varepsilon_\eu-\varepsilon_\ed)\alpha^2_\mathrm{cs}\slash(4\sqrt{\Gamma_1+\varepsilon_\ed})$. Indeed, the Fourier-spectrum shown in Fig.~\ref{fig:Facilitation}(b) shows that the frequency peak on the right, associated with the fast coherent transport, reappears [cf. with Fig.~\ref{fig:Blockage}(c) from the main text] and the exciton transport is facilitated.

%%% FACILITATION
\figFacilitation

%%% SECTION
\section{Analytical solution for minimal model}
\label{app:AnalyticalSolutionForMinimalModel}

Here, we analytically calculate the time evolution of the exciton density $\rho_3$ governed by the effective Hamiltonian \eqref{eq:MinimalModel} from the main text. The first step is to use the $\mathbb{Z}_2$-symmetry \cite{cotton1991} of Eq.~\eqref{eq:MinimalModel} from the main text to split the eigenstates into states transforming with respect to the \emph{gerade} and the \emph{ungerade} irreducible representation of $\mathbb{Z}_2$. This amounts to transforming Eq.~\eqref{eq:MinimalModel} from the main text to the basis $B=(\ket*{2},\ket*{\pm})$, $\ket*{\pm}=(\ket*{1}\pm\ket*{3})\slash\sqrt{2}$, where the first two states are \emph{gerade} and the last one is \emph{ungerade}. Using the associated basis-change rotation $R_B$, we arrive at
\begin{equation}
    R^\dagger_B H R_B
    =
    J
    \mqty[\delta & \sqrt{2} & \\
          \sqrt{2} & 0 & \\
          & & 0]
\,,
\end{equation}
where the matrix structure corresponds to the basis $(\ket*{2},\ket*{\pm})$ and we defined $\delta=\alpha^2_\mathrm{cs}(\varepsilon_\eu-\varepsilon_\ed)\slash\big(4J\sqrt{\Gamma^2_1+\varepsilon_\ed}\big)$. Due to the block structure, the problem is reduced to diagonalising a symmetric two-dimensional matrix. The eigenvalues are $J\lambda_\pm=J(\delta\pm\sqrt{\delta^2+8})\slash{2}$ and the eigenvectors define the rotation
\begin{equation}
    R'
    =
    \mqty[\frac{\sqrt{2}}{\sqrt{2+\lambda^2_-}} & \frac{-\sqrt{2}}{\sqrt{2+\lambda^2_+}} \\
          \frac{-\lambda_-}{\sqrt{2+\lambda^2_-}} & \frac{\lambda_+}{\sqrt{2+\lambda^2_+}} ]
\,.
\end{equation}
The eigenstates of the Hamiltonian in the original basis $(\ket*{1},\ket*{2},\ket*{3})$ are therefore given by the columns of the matrix $R=R_B(R'\oplus{1})$. With the eigenstates at hand, analytically calculating the time-evolution operator is straightforward. In particular, for the initial state $\ket*{1}$, i.e., the exciton is localised on ion 1, we find for the exciton density at ion 3 the expression
\begin{multline}
    \rho_3(t)
    =
    \frac{48+14\delta^2+\delta^4}{2(8+\delta^2)^2}
    +
    \frac{\cos(\sqrt{8+\delta^2}Jt)}{8+\delta^2}
\\
    -
    \frac{\delta}{2\sqrt{8+\delta^2}}
    \sin(\frac{\delta}{2}Jt)
    \sin(\frac{\sqrt{8+\delta^2}}{2}Jt)
\\
    -
    \frac{1}{2}
    \cos(\frac{\delta}{2}Jt)
    \cos(\frac{\sqrt{8+\delta^2}}{2}Jt)
\,.
\end{multline}
Expanding this expression to first order in small $\delta$, i.e., weak initial `kicks', yields:
\begin{equation}\label{eq:ApproximatedP3SmallDelta}
    \rho_3(t)
    \approx
    \frac{3}{8}
    +
    \frac{1}{8}\cos(\sqrt{8}Jt)
    -
    \frac{1}{2}\cos(\frac{\sqrt{8}}{2}Jt)
\,.
\end{equation}
The frequency $\sqrt{8}J$ corresponds to the right-most peak in Fig.~\ref{fig:Blockage}(c) from the main text and $\sqrt{8}J\slash{2}$ to the centre between the two lower-frequency ones. We numerically verified that the splitting into two frequency branches is caused by next-to-nearest-neighbour coupling, which we ignored in our simple model.

\end{document}